\documentclass[showpacs,pra,twocolumn]{revtex4}
\usepackage{graphicx}
\usepackage{amsmath}
\usepackage{amsfonts}
\usepackage{amssymb}
\usepackage{amsthm}
\newcommand{\ket}[1]{|#1\rangle}
\newcommand{\bra}[1]{\langle#1|}
\newcommand{\Real}{{\textrm{Re}}}

\newcommand{\Tr}{{\textrm{Tr}}} 
\newcommand{\Sp}{{\textrm{Sp}}}
\begin{document}
\title{Holonomy for Quantum Channels}
\author{David Kult$^{1}$\footnote{Electronic address: 
david.kult@kvac.uu.se}, Johan \AA berg$^{2}$\footnote{
Electronic address: J.Aberg@damtp.cam.ac.uk}, and Erik 
Sj\"oqvist$^{1}$\footnote{Electronic address: eriks@kvac.uu.se}} 
\affiliation{$^{1}$Department of Quantum Chemistry, 
Uppsala University, Box 518, Se-751 20 Uppsala, Sweden \\
$^{2}$Centre for Quantum Computation, Department of Applied
Mathematics and Theoretical Physics, University of Cambridge, 
Wilberforce Road, Cambridge CB3 0WA, United Kingdom}
\date{\today}
\begin{abstract}
A quantum holonomy reflects the curvature of some underlying structure
of quantum mechanical systems, such as that associated with quantum
states. Here, we extend the notion of holonomy to families of quantum
channels, i.e., trace-preserving completely positive maps. By the use
of the Jamio{\l}kowski isomorphism, we show that the proposed channel
holonomy is related to the Uhlmann holonomy. The general theory is
illustrated for specific examples. We put forward a physical
realization of the channel holonomy in terms of interferometry. This
enables us to identify a gauge-invariant physical object that directly
relates to the channel holonomy. Parallel transport condition and
concomitant gauge structure are delineated in the case of smoothly
parametrized families of channels. Finally, we point out that
interferometer tests that have been carried out in the past to confirm
the $4\pi$ rotation symmetry of the neutron spin, can be viewed as
early experimental realizations of the channel holonomy.
\end{abstract}
\maketitle
\section{Introduction} 
Quantum geometric phases and quantum holonomies have, since 
their initial discovery, proven to be a versatile structure
that appears in many different contexts in quantum mechanics.
Berry showed in his seminal paper \cite{berry84} that a state vector
initially in an eigenspace of a nondegenerate Hamiltonian acquires 
a geometric phase factor in addition to the familiar dynamical phase
factor after being adiabatically transported around a closed curve in
the parameter space of the Hamiltonian. Wilczek and Zee
\cite{wilczek84} soon thereafter extended Berry's work by showing that
the geometric phase factor generalizes to a unitary state change,
called a non-Abelian holonomy (or just holonomy for short), in the
case of degenerate Hamiltonians. Another extension of Berry's work
was provided by Aharonov and Anandan \cite{aharonov87} who removed the
requirement of adiabaticity by showing that a geometric phase factor
is defined for any cyclic evolution of a pure quantum state.  This
result was further generalized by Samuel and Bhandari \cite{samuel88} 
to include noncyclic evolution as well. A holonomy for curves of 
density operators was first introduced by Uhlmann \cite{uhlmann86}. 

The aforementioned geometric phases and holonomies may be classified
in the following way. Firstly, we have holonomies for subspaces, such as
eigenspaces of Hamiltonians \cite{berry84,wilczek84}, subspaces
selected by projective measurements \cite{anandan89}, cyclic
subspaces \cite{anandan88}, and decoherence free subspaces 
\cite{wu05,carollo06}. Holonomic quantum computation
\cite{zanardi99,pachos00} is related to this class of
holonomies. Secondly, we have geometric phases and holonomies for
quantum states, both pure \cite{aharonov87,samuel88} and mixed
\cite{uhlmann86}. The geometric structures of these two classes are
given by the fiber bundles associated with the mappings, ``basis of
subspace'' $\mapsto$ ``subspace'' and ``purification'' $\mapsto$ 
``state'', respectively.

In this paper, we focus on the geometry related to a third major
concept in quantum theory, namely quantum maps. More precisely, we are
interested in the holonomy for sequences of quantum
channels, i.e., trace-preserving completely positive maps.  This
concept, which we shall call ``channel holonomy'', is associated with
the geometry given by a fiber bundle structure related to the
mapping ``Kraus representation'' $\mapsto$ ``completely positive
map''. The aim with this analysis is to delineate this structure and to
examine its physical relevance.

The outline of this paper is as follows. In the next section, we
introduce the channel holonomy and examine its behavior under gauge
transformations. The relation between the Uhlmann holonomy
\cite{uhlmann86} for a sequence of density operators, constructed 
from the Jamio{\l}kowski isomorphism \cite{jamiolkowski72}, and the
channel holonomy is analyzed in Sec. \ref{sec:jamiolkowski}. The case
of smoothly parametrized families of quantum channels is discussed in
Sec. \ref{sec:smooth}. We derive the parallel transport condition and
introduce a gauge potential associated with such families. In
Sec. \ref{sec:examples}, the channel holonomy is calculated for
specific types of channel sequences. A physical realization of the
channel holonomy based on ancillary constructions in two-beam
interferometry is demonstrated in Sec. \ref{sec:realization}. In
particular, we demonstrate that the channel holonomy is related to the
``gluing matrix'' \cite{aberg04a,aberg04b} that arises when two channels
are combined in an interferometer.  The interferometer setup also
provides means to identify a physically meaningful gauge-invariant
object associated with the channel holonomy. In Sec. 
\ref{sec:realizationsmooth}, we examine the case of smoothly 
parametrized families of quantum channels using the ancillary
construction. It should be noted that the analysis in
Secs. \ref{sec:realization} and \ref{sec:realizationsmooth} parallel
to a large extent that of Secs. \ref{sec:chaho} and \ref{sec:smooth},
the main difference being that while the latter utilizes directly the
Kraus operators, the former utilizes ancillary systems. In
Sec. \ref{sec:4pi}, we examine the interferometer tests in
Refs. \cite{rauch75,werner75} of the $4\pi$ symmetry of the neutron
spin in terms of the channel holonomy. The paper ends with the
conclusions.

\section{Channel holonomy}
\label{sec:chaho}
Consider a trace-preserving completely positive map (channel for
short) $\mathcal{F}$ acting on a $D$-dimensional state space of a
quantum system.  The action of the channel on a state $\rho$ can be
expressed as
\begin{eqnarray}
\mathcal{F}(\rho) & = & \sum_{k=1}^K F_k\rho F_k^{\dagger}.
\end{eqnarray}
The operators $F_k$ constitute a Kraus representation
of $\mathcal{F}$ \cite{kraus83}. We assume that $K$ is the number 
of linearly independent Kraus operators needed to represent 
$\mathcal{F}$, i.e., $K$ is the Kraus number $K(\mathcal{F})$
of the channel \cite{aberg04a}. From trace preservation it follows
that $\sum_k F_k^\dagger F_k = \hat{1}$, where $\hat{1}$ is the identity
operator. The Kraus representation of a channel is not unique. If
$\{F_{k}\}_k$ is a valid representation of $\mathcal{F}$ then so is
$\{\widetilde{F}_k\}_k$, where
\begin{equation}
\label{eq:kraustransform}
\widetilde{F}_k = \sum_{l=1}^K  F_l \boldsymbol{U}_{lk} , 
\end{equation}
$\boldsymbol{U}$ being a unitary matrix \cite{remark1}. 

Let $\{ E_k \}_k$ be a linearly independent Kraus representation of a
channel $\mathcal{E}$. Given another channel $\mathcal{F}$ with
$K(\mathcal{F})=K(\mathcal{E}) = K$, we wish to find a linearly
independent Kraus representation $\{ F_k \}_k$ of $\mathcal{F}$ that
in some sense is parallel with $\{ E_k \}_k$. A
convenient choice would be to find the Kraus representation $\{ F_k
\}_k$ that minimizes \cite{remark2}
\begin{eqnarray}
\label{maxparall}
\sum_k\| E_k-F_k\|^2 = 2D - 
2 \Real \Tr  \boldsymbol{T} ,
\end{eqnarray}
where $\|\cdot \|$ denotes the Hilbert-Schmidt norm and 
$\boldsymbol{T}$ is a matrix with elements $\boldsymbol{T}_{kl} = 
\Tr ( F_k^{\dagger} E_l)$. Under a change of Kraus 
representation $\{ F_k \}_k \rightarrow \{\widetilde{F}_k\}_k$, 
as given by Eq. (\ref{eq:kraustransform}), the matrix $\boldsymbol{T}$ 
transforms as $\boldsymbol{T} \rightarrow \widetilde{\boldsymbol{T}}= 
\boldsymbol{U}^{\dagger} \boldsymbol{T}$. Hence, 
\begin{eqnarray}
\sum_k\| E_k-\widetilde{F}_k\|^2 = 2D  - 
2 \Real \Tr (\boldsymbol{U}^{\dagger} \boldsymbol{T}) .
\end{eqnarray}
Assuming that $\boldsymbol{T}$ is of rank $K$ (see the appendix for an 
elaboration on the rank of $\boldsymbol{T}$) the minimum is obtained 
when $\widetilde{\boldsymbol{T}} > 0$ which corresponds to the 
choice $\boldsymbol{U} = \Phi(\boldsymbol{T}) \equiv \sqrt{\boldsymbol{T} 
\boldsymbol{T}^{\dagger}}^{-1}\boldsymbol{T}$. Thus, we say that two Kraus 
representations $\{ E_k \}_k$ and $\{ F_k \}_k$ are parallel if their
corresponding $\boldsymbol{T}$ matrix is positive definite.

We are now in a position where we can define a holonomy corresponding
to a sequence $\mathcal{E}_1, \ldots,\mathcal{E}_N$ of quantum
channels with $K(\mathcal{E}_1)=\ldots=K(\mathcal{E}_N) = K$. This is
done by choosing a Kraus representation $\{ E_k^n \}_{k=1}^K$ for each
$\mathcal{E}_n$ in the sequence and encoding the Kraus freedom in a
family of unitary matrices $\boldsymbol{U}_n$, $n = 1,\ldots,N$. In
this way, we may express the parallelity conditions as
\begin{eqnarray}
\boldsymbol{U}_{n+1}^{\dagger} \boldsymbol{T}_{n+1,n} 
\boldsymbol{U}_n > 0, \quad n = 1,\ldots,N-1, 
\label{eq:chparallel}
\end{eqnarray}
where the $K\times K$ matrices $\boldsymbol{T}_{n+1,n}$ with 
elements 
\begin{equation}
[\boldsymbol{T}_{n+1,n}]_{kl} = \Tr ( {E_k^{n+1}}^{\dagger} 
E_l^n)
\end{equation}
are all assumed to be of rank $K$. The conditions 
in Eq. (\ref{eq:chparallel}) are satisfied by 
\begin{eqnarray}
\boldsymbol{U}_{n+1} = \Phi( \boldsymbol{T}_{n+1,n}) 
\boldsymbol{U}_n .
\end{eqnarray}
We obtain after iteration
\begin{eqnarray}
\boldsymbol{U}_{N} = \Phi(\boldsymbol{T}_{N,N-1}) \cdots 
\Phi(\boldsymbol{T}_{2,1}) \boldsymbol{U}_1. 
\end{eqnarray}
Define 
\begin{eqnarray}
\boldsymbol{U}_{\textrm{\textrm{ch}}} 
(\mathcal{E}_1,\ldots,\mathcal{E}_N) = \Phi(\boldsymbol{T}_{1,N}) 
\boldsymbol{U}_N \boldsymbol{U}_1^{\dagger} 
\nonumber \\ 
 = \Phi(\boldsymbol{T}_{1,N}) 
\Phi(\boldsymbol{T}_{N,N-1}) \cdots \Phi(\boldsymbol{T}_{2,1})  
\label{eq:kinholonomy}
\end{eqnarray}
to be the channel holonomy for the sequence 
$\mathcal{E}_1,\ldots,\mathcal{E}_N$.

If we consider the set of linearly independent Kraus representations
as a fiber bundle with the set of channels with a fixed Kraus number
as base manifold, the change of Kraus representations, as in
Eq. (\ref{eq:kraustransform}), can be interpreted as a gauge
transformation. As seen from the definition in
Eq. (\ref{eq:kinholonomy}) the factor $\boldsymbol{U}_N
\boldsymbol{U}_1^{\dagger}$ is multiplied from the left by 
$\Phi[\boldsymbol{T}_{1,N}]$. This construction guarantees that the
matrix
$\boldsymbol{U}_{\textrm{\textrm{ch}}}[\mathcal{E}_1,\ldots,\mathcal{E}_N]$
transforms gauge covariantly.  To see this we make the gauge
transformation
\begin{equation}
\label{gaugetransf}
E_k^n \rightarrow \sum_l 
E_l^n  \left[ \boldsymbol{V}_n \right]_{lk}
\end{equation}
of the Kraus representation 
of each channel in the sequence. We obtain 
\begin{eqnarray} 
\Phi (\boldsymbol{T}_{n+1,n}) \rightarrow  
\boldsymbol{V}_{n+1}^{\dagger} \Phi (\boldsymbol{T}_{n+1,n}) 
\boldsymbol{V}_n , 
\end{eqnarray}
which implies 
\begin{eqnarray} 
\label{covariant}
\boldsymbol{U}_{\textrm{\textrm{ch}}} (\mathcal{E}_1,\ldots,\mathcal{E}_N) 
\rightarrow \boldsymbol{V}_1^{\dagger} 
\boldsymbol{U}_{\textrm{\textrm{ch}}}(\mathcal{E}_1,\ldots,\mathcal{E}_N) 
\boldsymbol{V}_1 . 
\end{eqnarray}
Hence, the channel holonomy transforms gauge covariantly as required. 

Concerning the gauge covariance, let us point out that the channel 
holonomy is described as a unitary matrix. The gauge
covariance of this matrix is necessary since it, in some sense, is a
matrix representation of a gauge-invariant object, and the gauge
covariance reflects the freedom of the choice of ``basis'' in this
matrix representation. It is to be noted \cite{comment} that this is
the case also for other type of holonomies, such as those of
Refs. \cite{wilczek84,kult06}. In Sec. \ref{sec:realization} 
we elucidate what gauge-invariant object the channel holonomy 
represents, and in what sense the channel holonomy is a matrix 
representation of this object.

An alternative way to obtain the channel holonomy defined above is 
to make a gauge transformation yielding $\boldsymbol{T}_{n+1,n} 
\rightarrow \widetilde{\boldsymbol{T}}_{n+1,n} >0$, for all 
$n=1,\ldots,N-1$, i.e., $\Phi (\widetilde{\boldsymbol{T}}_{n+1,n}) = 
\boldsymbol{I}$, $\boldsymbol{I}$ being the $K\times K$ identity 
matrix. This choice amounts to parallel transport along the 
sequence $\mathcal{E}_1,\ldots,\mathcal{E}_N$. From Eq. 
(\ref{eq:kinholonomy}) it follows that   
\begin{eqnarray}
\boldsymbol{U}_{\textrm{\textrm{ch}}} 
(\mathcal{E}_1,\ldots,\mathcal{E}_N) = 
\Phi(\widetilde{\boldsymbol{T}}_{1,N}) , 
\label{eq:ptholonomy}
\end{eqnarray}
which is the expression for the channel holonomy in the parallel 
transport gauge. 

\section{Relation to the Uhlmann holonomy}
\label{sec:jamiolkowski} 
For any given channel $\mathcal{E}$ acting on elements of
$\mathcal{L}(\mathcal{H}_q)$, i.e., the set of linear operators on the
$D$-dimensional Hilbert space $\mathcal{H}_q$, one can find a
corresponding density operator $\rho \in
\mathcal{L}(\mathcal{H}_q\otimes \mathcal{H}_q)$. This can be 
done via the Jamio{\l}kowski isomorphism
\cite{jamiolkowski72}, i.e., $\mathcal{E} \mapsto \rho = 
\mathcal{E} \otimes \mathcal{I} (|\psi \rangle \langle \psi |)$,
where $\mathcal{I}$ is the identity channel and 
$|\psi \rangle = \frac{1}{\sqrt{D}}\sum_{k=1}^{D} |k \rangle \otimes
|k\rangle $ with $\{|k \rangle \}_k$ an orthonormal basis of
$\mathcal{H}_q$. We show that the holonomy associated with  
a sequence of channels $\mathcal{E}_1, \ldots ,\mathcal{E}_N$ is 
related to the Uhlmann holonomy \cite{uhlmann86} for the 
extended sequence of density operators $\rho_1, \ldots, \rho_N, 
\rho_{N+1} = \rho_1$ \cite{remark3}, where 
\begin{equation}
\label{jamiol}
\rho_n = \mathcal{E}_n \otimes \mathcal{I} 
(|\psi \rangle \langle \psi |) . 
\end{equation} 
To avoid technical complications, we assume that all channels have
maximal Kraus number $K(\mathcal{E}_n) = D^2$. This guarantees that
the corresponding density operators $\rho_n$ are faithful
\cite{uhlmann86}, i.e., they are full rank.

There are several ways to calculate the Uhlmann holonomy associated
with a sequence $\rho_1,\ldots,\rho_{N+1}$ of faithful density
operators. For each density operator $\rho_n$ there corresponds
Uhlmann amplitudes $\widetilde{W}_n =
\sqrt{\rho_n}V_n$, where $V_n = \Phi(\widetilde{W}_n) =
\sqrt{\widetilde{W}_n\widetilde{W}_n^{\dagger}}^{-1}\widetilde{W}_n$ 
is a unitary operator. A sequence of such amplitudes is parallel if
$\widetilde{W}_{n+1}^{\dagger}\widetilde{W}_n>0$, which allows us to 
define the Uhlmann holonomy $U_{\textrm{Uhl}} = V_{N+1}V_{1}^{\dagger}$. 
If we instead consider another sequence of amplitudes $W_n$ that is not
parallel transported (but corresponds to the same sequence of density
operators), we can make it into a parallel transported sequence
$W_nU_n$ by a choice of unitary operators $U_n$ such that
\begin{equation}
U_{n+1}^{\dagger}W_{n+1}^{\dagger}W_n U_n > 0 , 
\end{equation}
which implies that $U_{n+1}=\Phi (W_{n+1}^{\dagger}W_n) U_n$ and 
we find $U_{\textrm{Uhl}} = \Phi(W_{N+1})U_{N+1}U_{1}^{\dagger} 
\Phi(W_1^{\dagger})$. If we furthermore assume that the sequence 
of amplitudes is cyclic, i.e., $W_{N+1} = W_{1}$ (which implies 
that the underlying sequence of density operators is cyclic, 
i.e., $\rho_{N+1} = \rho_{1}$), we obtain 
\begin{equation}
U_{\textrm{Uhl}} = 
\Phi(W_{1})U_{N+1}U_{1}^{\dagger}\Phi(W_1^{\dagger}).
\end{equation} 
We also note that 
\begin{equation}
\label{UN1}
U_{N+1} = \Phi(W_{1}^{\dagger}W_{N})\Phi(W_{N}^{\dagger}W_{N-1}) 
\cdots \Phi(W_{2}^{\dagger}W_{1})U_{1}.
\end{equation}

Let us now return to the cyclic sequence of density operators 
$\rho_{1},\ldots,\rho_{N},\rho_{N+1}=\rho_{1}$ as defined by Eq. 
(\ref{jamiol}). We fix an orthonormal basis $\{|f_k \rangle \}_k$ 
of $\mathcal{H}_q \otimes \mathcal{H}_q$, and define the amplitudes
\begin{eqnarray}
\label{eq:amplitude}
W_n = \sum_k (E_k^n \otimes \hat{1})|\psi \rangle \langle f_k |,
\end{eqnarray}
where $\{ E_k^n \}_{k}$ is a linearly independent Kraus representation
of $\mathcal{E}_n$. We furthermore find that
\begin{eqnarray}
W_{n+1}^\dagger W_n & = & \frac{1}{D}\sum_{kl} |f_k \rangle \langle f_l| 
\Tr({E_k^{n+1}}^{\dagger}E_l^n) 
\nonumber \\ 
 & = & \frac{1}{D}\sum_{kl} |f_k \rangle \langle f_l| 
[\boldsymbol{T}_{n+1,n}]_{kl} .
\end{eqnarray}
It follows that the two amplitudes $W_n$ and $W_{n+1}$ defined 
in Eq. (\ref{eq:amplitude}) are parallel if and only 
if $\boldsymbol{T}_{n+1,n} >0$. Hence, the parallelity condition for 
channels is, via this construction, closely related to the Uhlmann 
parallel transport of amplitudes. If we combine Eq. (\ref{UN1}) with 
the fact that $\Phi(W_{n+1}^{\dagger}W_n) =  
\sum_{kl} |f_{k}\rangle\langle f_{l}| [\Phi(\boldsymbol{T}_{n+1,n})]_{kl}$ 
we can conclude that
\begin{equation}
U_{N+1} = \sum_{kl}|f_k \rangle \langle f_l| 
[\boldsymbol{U}_{\textrm{ch}}]_{kl}U_{1} , 
\end{equation}
which implies
\begin{equation}
\label{relation}
[\boldsymbol{U}_{\textrm{ch}}]_{kl} = 
\langle f_{k}|\Phi(W_1^{\dagger})  
U_{\textrm{Uhl}}\Phi(W_{1})|f_{l}\rangle . 
\end{equation}
The right-hand side of the above equation transforms as in
Eq. (\ref{covariant}) under the gauge transformation in
Eq. (\ref{gaugetransf}), as it should. We note that it is
$\Phi(W_{1})$ and $\Phi(W_1^{\dagger})$ that are ``responsible'' for
the gauge covariance, as $U_{\textrm{Uhl}}$ remains invariant. Note
also that the construction in Eq. (\ref{eq:amplitude}) leading to
Eq. (\ref{relation}) contains an arbitrary choice of basis
$\{|f_{k}\rangle\}_{k}$, as well as an arbitrary maximally entangled
state $|\psi\rangle$. The channel holonomy
$\boldsymbol{U}_{\textrm{ch}}$ should not depend on any of these
arbitrary choices. (This invariance of $\boldsymbol{U}_{\textrm{ch}}$
with respect to the choice of $\{|f_{k}\rangle\}_{k}$ and
$|\psi\rangle$ should not be confused with the covariance of
$\boldsymbol{U}_{\textrm{ch}}$ under the gauge transformations in
Eq.~(\ref{gaugetransf}).)  The Uhlmann holonomy $U_{\textrm{Uhl}}$ is
invariant under change of $\{|f_{k}\rangle\}_{k}$, and although
$\Phi(W_{1})$ depends on $\{|f_{k}\rangle\}_{k}$, the combination
$\Phi(W_{1})|f_{k}\rangle$ does not, which leaves
$\boldsymbol{U}_{\textrm{ch}}$ invariant under the choice of
$\{|f_{k}\rangle\}_{k}$. Given a maximally entangled state
$|\psi\rangle$, all other maximally entangled states can be obtained
as $[\hat{1} \otimes U]|\psi\rangle$, where $U$ is unitary. Both
$U_{\textrm{Uhl}}$ and $\Phi(W_{1})$ depend on the choice of
$|\psi\rangle$. However, the combination
$\Phi(W_1^{\dagger})U_{\textrm{Uhl}}\Phi(W_{1})$ is independent of the
choice of maximally entangled state $|\psi\rangle$. We can thus
conclude that $\boldsymbol{U}_{\textrm{ch}}$ behaves as required.

\section{Smoothly parametrized families of channels}
\label{sec:smooth} 
\subsection{The parallel transport condition}
\label{sec:smoothparallel}
So far we have considered the holonomies associated with sequences of
channels. Here we consider the transition to families of channels
$\mathcal{E}_s$ smoothly parameterized by a real variable
$s\in[0,1]$. As before we assume that the Kraus number is constant
within each family, i.e., $K( \mathcal{E}_s ) = K$ for all
$s$. Consider a smoothly parametrized family of linearly independent
Kraus representations $\{E_{k}(s)\}_{k=1}^{K}$ of $\mathcal{E}_s$. For
sequences of Kraus representations, the parallel transport condition
is $\boldsymbol{T}_{n+1,n}>0$. A generalization of this condition to
smooth curves is obtained by requiring that the matrix with elements
$\Tr[E_{k}^{\dagger}(s+\delta s)E_{l}(s)]$ is positive definite to
first order in the limit of small $\delta s$. We find
\begin{equation}
\Tr[E_{k}^{\dagger}(s+\delta s)E_{l}(s)] = 
\boldsymbol{Q}_{kl} + \delta s \boldsymbol{R}_{kl},
\end{equation}
where 
\begin{equation}
\label{eq:QRdef}
\boldsymbol{Q}_{kl} = \Tr[E_k^{\dagger}(s)E_{l}(s)], \quad 
\boldsymbol{R}_{kl} = \Tr[\dot{E}_k^{\dagger}(s)E_{l}(s)].
\end{equation}
Since $\{ E_k (s) \}_k$ is a linearly independent set it follows that
$\boldsymbol{Q}$ is positive definite. Therefore, a necessary
condition for $\boldsymbol{Q} + \delta s\boldsymbol{R}$ to be positive
definite is that $\boldsymbol{R}$ is Hermitian. We shall now see that
this is also a sufficient condition for small $\delta s$. Since
$\boldsymbol{Q}$ is positive definite it follows that
\begin{equation}
\boldsymbol{Q}+ \delta s\boldsymbol{R} = 
\sqrt{\boldsymbol{Q}}(\boldsymbol{I} + 
\delta s \sqrt{\boldsymbol{Q}}^{-1} \boldsymbol{R} 
\sqrt{\boldsymbol{Q}}^{-1})\sqrt{\boldsymbol{Q}}.
\end{equation}
The assumption $\boldsymbol{R}^{\dagger}=\boldsymbol{R}$ implies  
that $\sqrt{\boldsymbol{Q}}^{-1}\boldsymbol{R}\sqrt{\boldsymbol{Q}}^{-1}$
is a finite Hermitian matrix and consequently its eigenvalues are real
and $\boldsymbol{Q}+ \delta s\boldsymbol{R}$ is positive definite for 
sufficiently small $\delta s$. We can conclude that a smooth 
family of Kraus representations $\{E_{k}(s)\}_{k=1}^{K}$ of channels 
with constant Kraus number $K$ is parallel transported if and only if
\begin{equation}
\label{eq:smoothparallel}
\Tr[\dot{E}_{k}^{\dagger}(s)E_{l}(s)] = 
\Tr[E_{k}^{\dagger}(s)\dot{E}_{l}(s)],\quad k,l = 1,\ldots, K.
\end{equation}
This can be compared with the Uhlmann parallel transport condition for
smoothly parameterized families of amplitudes $W(s)$, which is
$\dot{W}^{\dagger}(s)W(s) = W^{\dagger}(s)\dot{W}(s)$
\cite{uhlmann86}.

\subsection{The gauge potential}
We now demonstrate how to introduce a gauge potential along the path 
$C: [0,1] \ni s \mapsto \mathcal{E}_s$ of smoothly parametrized 
channels. Suppose we choose a family of Kraus representations 
$\{E_{k}(s)\}_{k=1}^{K}$ over $C$ that is not parallel transported. 
We can make a gauge transformation of the form 
\begin{eqnarray}
E_k (s) \rightarrow \widetilde{E}_k (s) = 
\sum_{l=1}^K E_{l}(s) \boldsymbol{U}_{lk} (s) 
\label{eq:smoothgt}
\end{eqnarray}
such that $\{\widetilde{E}_{k}(s)\}_{k=1}^{K}$ is parallel transported. 
By inserting Eq. (\ref{eq:smoothgt}) into Eq. (\ref{eq:smoothparallel}) 
we obtain  
\begin{eqnarray} 
\boldsymbol{R} - \boldsymbol{R}^{\dagger} = 
\dot{\boldsymbol{U}} \boldsymbol{U}^{\dagger} \boldsymbol{Q} +   
\boldsymbol{Q} \dot{\boldsymbol{U}} \boldsymbol{U}^{\dagger} , 
\label{eq:potential1}
\end{eqnarray}
where $\boldsymbol{Q}$ and $\boldsymbol{R}$ are as in Eq. 
(\ref{eq:QRdef}). Let $\boldsymbol{A}(s)$ be an anti-Hermitian
matrix generating $\boldsymbol{U} (s)$ via $\dot{\boldsymbol{U}} =
\boldsymbol{A} \boldsymbol{U}$, and substitute into Eq. 
(\ref{eq:potential1}) yielding (cf. Ref. \cite{uhlmann91})
\begin{eqnarray} 
\boldsymbol{R} - \boldsymbol{R}^{\dagger} = 
\boldsymbol{A} \boldsymbol{Q} + \boldsymbol{Q} \boldsymbol{A}. 
\label{eq:potential2}
\end{eqnarray}
If we assume $\boldsymbol{Q}(s)>0$ for all $s$, then, according to
Theorem VII.2.3 in Ref. \cite{bhatia97}, we find that
Eq. (\ref{eq:potential2}) has a unique solution $\boldsymbol{A}(s)$
for each $s$, namely
\begin{equation}
\boldsymbol{A}(s) = \int_{0}^{\infty} e^{-r\boldsymbol{Q}(s)} 
[\boldsymbol{R}(s)-\boldsymbol{R}^{\dagger}(s)] 
e^{-r\boldsymbol{Q}(s)}dr.
\end{equation}
One can see from the right-hand side of this equation that 
$\boldsymbol{A} (s)$ is indeed an anti-Hermitian matrix for 
all $s\in [0,1]$.

To see that $\boldsymbol{A}$ transforms as a proper gauge potential
\cite{yang54} we consider an arbitrary gauge transformation
$E_{k}(s)\mapsto \sum_{l}E_{l}(s)\boldsymbol{V}_{lk}(s)$, where
$\boldsymbol{V}(s)$ is a smooth family of unitary operators. We find
that $\boldsymbol{Q} \rightarrow \boldsymbol{V}^{\dagger}
\boldsymbol{Q}\boldsymbol{V}$ and $\boldsymbol{R} \rightarrow 
\dot{\boldsymbol{V}}^{\dagger} \boldsymbol{Q} \boldsymbol{V} + 
\boldsymbol{V}^{\dagger} \boldsymbol{R} \boldsymbol{V}$.
From  these transformation properties we obtain 
\begin{equation} 
\boldsymbol{A} \rightarrow \boldsymbol{V}^{\dagger} \boldsymbol{A} 
\boldsymbol{V} +  \dot{\boldsymbol{V}}^{\dagger} \boldsymbol{V} , 
\end{equation}
i.e., $\boldsymbol{A}$ indeed transforms as a gauge potential. 

\subsection{An example}
The gauge potential $\boldsymbol{A} (s)$ along $C$ can be found by 
solving Eq. (\ref{eq:potential2}). Although this is difficult in the general 
case, it can be done for Kraus number $K=2$. 
Let $\boldsymbol{\sigma} = (\boldsymbol{\sigma}_x,\boldsymbol{\sigma}_y,
\boldsymbol{\sigma}_z)$ be the standard
$2\times 2$ Pauli matrices. Since $\boldsymbol{Q} >0$ and $\Tr
\boldsymbol{Q} =
\Tr \boldsymbol{I}$, we may write 
\begin{eqnarray}
\boldsymbol{Q} = \boldsymbol{I} + {\bf x} \cdot \boldsymbol{\sigma}, 
\label{eq:q}
\end{eqnarray} 
where $|{\bf x}| \neq 1$ (the eigenvalues of $\boldsymbol{Q}>0$ 
are $1\pm |{\bf x}|$). Furthermore, from the definition 
of the matrices $\boldsymbol{Q}$ and $\boldsymbol{R}$, it follows 
that $\Tr \boldsymbol{R} + \Tr \boldsymbol{R}^{\dagger} = 
\Tr \dot{\boldsymbol{Q}} = 0$, which implies  
$\Real \Tr \boldsymbol{R} =0$. Thus, we may put 
\begin{eqnarray}
\boldsymbol{R} = iz_0 \boldsymbol{I} + ({\bf y} + i{\bf z}) \cdot 
\boldsymbol{\sigma} ,   
\label{eq:r}
\end{eqnarray} 
where $2 {\bf y} = \dot{{\bf x}}$. 
Finally, $\boldsymbol{A}$ is anti-Hermitian, which suggests that we
can write 
\begin{eqnarray}
\boldsymbol{A} = 
i u_0 \boldsymbol{I} + i {\bf u} \cdot \boldsymbol{\sigma} . 
\label{eq:a}
\end{eqnarray}
We assume that ${\bf x},z_0,{\bf y},{\bf z},u_0,{\bf u}$ are all 
smooth real-valued functions of $s$. By inserting Eqs. (\ref{eq:q}), 
(\ref{eq:r}), and (\ref{eq:a}) into Eq. (\ref{eq:potential2}), we 
obtain 
\begin{eqnarray}
\boldsymbol{A} & = & 
i \left( \frac{z_0 - {\bf x} \cdot {\bf z}}{1-|{\bf x}|^2} \right) 
\boldsymbol{I} 
\nonumber \\ 
 & & + i\left( \frac{{\bf z} - z_0 {\bf x} + {\bf x} \times ({\bf x} 
\times {\bf z}) }{1-|{\bf x}|^2} \right) \cdot \boldsymbol{\sigma}  
\label{eq:aqubit}
\end{eqnarray}
along $C$, which is well-defined since $|{\bf x}| \neq 1$. Note, in
particular, that $\boldsymbol{A}$ vanishes when $z_0$ and ${\bf z}$
vanish, which from Eq. (\ref{eq:r}) can be seen to correspond to the
parallel transport condition $\boldsymbol{R}^{\dagger} =
\boldsymbol{R}$. Thus, the form of the $K=2$ gauge potential is
consistent with the notion of parallel transport developed in
Sec. \ref{sec:smoothparallel}.  Note also that a result analogous to
Eq. (\ref{eq:aqubit}) for the Uhlmann holonomy in the case of smooth
families of faithful qubit density operators has been obtained in
Ref. \cite{dittmann92}.

\subsection{Writing the holonomy in terms of the gauge potential}
It remains to demonstrate that the holonomy of the smooth path $C:
[0,1] \ni s \mapsto \mathcal{E}_s$ can be expressed in terms of
$\boldsymbol{A}$.  Consider the polar decomposition
$\boldsymbol{T}_{s+\delta s,s} = |\boldsymbol{T}_{s+\delta s,s}| \Phi
(\boldsymbol{T}_{s+\delta s,s})$. To the first order in $\delta s$ we
may write $|\boldsymbol{T}_{s+\delta s,s}| = \boldsymbol{Q} +
\delta s \boldsymbol{H}$ and $\Phi (\boldsymbol{T}_{s+\delta s,s}) = 
\boldsymbol{I} + \delta s \boldsymbol{J}$, where 
$\boldsymbol{H}^{\dagger} = \boldsymbol{H}$ and 
$\boldsymbol{J}^{\dagger} = -\boldsymbol{J}$. 
Thus, $\boldsymbol{T}_{s+\delta s,s} = \boldsymbol{Q} + \delta s 
(\boldsymbol{Q} \boldsymbol{J} + \boldsymbol{H} ) + O(\delta s^2)$
and we obtain 
\begin{eqnarray} 
\boldsymbol{T}_{s+\delta s,s} - 
\boldsymbol{T}_{s+\delta s,s}^{\dagger} & = & 
\delta s (\boldsymbol{Q} \boldsymbol{J} + 
\boldsymbol{J} \boldsymbol{Q}) + O(\delta s^2) . 
\end{eqnarray}
Furthermore, from Eq. (\ref{eq:potential2}) and the expression 
$\boldsymbol{T}_{s+\delta s,s} = \boldsymbol{Q} + \delta s 
\boldsymbol{R} + O(\delta s^2)$, we find 
\begin{eqnarray} 
\boldsymbol{T}_{s+\delta s,s} - 
\boldsymbol{T}_{s+\delta s,s}^{\dagger} = 
\delta s (\boldsymbol{Q} \boldsymbol{A} + 
\boldsymbol{A} \boldsymbol{Q}) + O(\delta s^2) . 
\end{eqnarray}
Since $\boldsymbol{Q} > 0$ it follows that 
$\boldsymbol{J} = \boldsymbol{A}$ and the channel holonomy can be 
written as 
\begin{eqnarray} 
\boldsymbol{U}_{{\textrm{ch}}} (C) = \Phi (\boldsymbol{T}_{0,1})
{\bf P} e^{\int_0^1 \boldsymbol{A} ds} ,
\end{eqnarray}
where ${\bf P}$ denotes path ordering and $\boldsymbol{A}$ is a 
solution of Eq. (\ref{eq:potential2}). 

\section{Examples}
\label{sec:examples}
In this section, we consider two examples where the channel holonomy
can be explicitly calculated. The first example concerns unitary channels,
i.e., operations on closed quantum systems. It turns out that the
channel holonomy for such channels is intimately connected to certain
cases of the mixed state phase proposed in Ref.
\cite{sjoqvist00}. Our second example concerns channels for which 
there exist Kraus representations that are built up by subspace 
holonomies associated with smooth paths that, e.g., can be 
approximately generated by sequential projective measurements 
\cite{kult06}. We call these ``holonomic channels''. As we 
demonstrate below, the eigenvalues of the channel holonomy for 
smooth families of holonomic channels are directly related to the  
trace of the subspace holonomies, which reduces to Wilson loops 
in the sense of Ref. \cite{zee88} for closed paths. 

\subsection{Unitary channels} 
Here, we provide a detailed analysis of the holonomy for sequences of
unitary channels, each of which acts on a $D$-dimensional state
space. Unitary channels are in a sense the simplest type of channels,
since their Kraus representations consist of a single unitary operator
($K=1$). Consequently, the Kraus freedom is encoded in a phase factor,
and it follows that the holonomy in Eq. (\ref{eq:kinholonomy}) is also
a phase factor, which we denote $\gamma_{\textrm{\textrm{ch}}}$. For a
sequence $\mathcal{E}_1^{{\textrm{u}}},\mathcal{E}_2^{{\textrm{u}}}, 
\ldots,\mathcal{E}_N^{{\textrm{u}}}$ of unitary 
channels represented by the unitary operators $U_1,U_2,\ldots,U_N$, 
the holonomy $\gamma_{\textrm{\textrm{ch}}}$ can be written
\begin{eqnarray}
\gamma_{\textrm{\textrm{ch}}} (\mathcal{E}_1^{{\textrm{u}}},\ldots,
\mathcal{E}_N^{{\textrm{u}}}) & = &  
\Phi \boldsymbol{(} \Tr ( U_1^{\dagger} U_N) \boldsymbol{)} 
\Phi \boldsymbol{(} \Tr ( U_N^{\dagger} U_{N-1} ) \boldsymbol{)} \ldots 
\nonumber \\ 
 & & \times \Phi \boldsymbol{(} \Tr ( U_2^{\dagger} U_1 ) \boldsymbol{)} , 
\label{eq:utwist}
\end{eqnarray}
where $\Phi(z)=z/|z|$ for any nonzero complex number $z$. Note that 
$\gamma_{\textrm{\textrm{ch}}} (\mathcal{E}_1^{{\textrm{u}}},\ldots,
\mathcal{E}_N^{{\textrm{u}}})$ 
is gauge invariant in the sense that it is unchanged under the gauge 
transformation $U_n \rightarrow e^{i\alpha_n} U_n$ for arbitrary 
real numbers $\alpha_n$, $n = 1,\ldots,N$. In particular, 
a gauge transformation $U_n \rightarrow \widetilde{U}_n$ such 
that 
\begin{equation}
\label{parallelcond}
\Phi \boldsymbol{(}\Tr ( \widetilde{U}_{n+1}^{\dagger} 
\widetilde{U}_n) \boldsymbol{)} >0
\end{equation}
yields
\begin{eqnarray}
\gamma_{\textrm{\textrm{ch}}} (\mathcal{E}_1^{{\textrm{u}}},\ldots,
\mathcal{E}_N^{{\textrm{u}}}) & = &  
\Phi \boldsymbol{(} \Tr ( \widetilde{U}_1^{\dagger} 
\widetilde{U}_N ) \boldsymbol{)}, 
\end{eqnarray} 
which is the channel holonomy in the parallel gauge (cf. Eq. 
(\ref{eq:ptholonomy}))

Consider now a smoothly parametrized family of unitary channels
$\mathcal{E}_s^{{\textrm{u}}} (\rho) = U(s)\rho U^{\dagger} (s)$, 
$s\in [0,1]$, and consider the equation of motion $i\dot{U}(s) = 
H(s)U(s)$, $H(s)$ being the ``Hamiltonian'' of the system ($\hbar = 1$). 
This family defines the path $C:[0,1]\ni s \mapsto 
\mathcal{E}_s^{{\textrm{u}}}$. To the first order in 
$\delta s$ we have $\Tr[U^{\dagger}(s+\delta s)U(s)] = D + 
i\delta s\Tr H(s)$, and $\Phi\boldsymbol{(} \Tr[ 
U^{\dagger}(s+\delta s)U(s)]\boldsymbol{)} = 
e^{\frac{i}{D}\delta s\Tr H(s)}$. In the $\delta s\rightarrow 0$ limit, 
Eq. (\ref{eq:utwist}) thus becomes 
\begin{eqnarray}   
\gamma_{\textrm{\textrm{ch}}} (C) & = & 
\Phi \boldsymbol{(} \Tr [ U^{\dagger}(0) U(1) ] \boldsymbol{)} 
\nonumber \\ 
 & & \times \exp \left[ \frac{i}{D} \int_0^1 ds \Tr H(s)  
\right] ,  
\label{eq:kinunitary}
\end{eqnarray}   
which resembles the Aharonov-Anandan geometric
phase \cite{aharonov87} with the ``dynamical phase'' $-\frac{1}{D}
\int_0^1 ds \Tr H(s)$ removed from the ``total phase'' 
$\arg\Phi \boldsymbol{(} \Tr [U^{\dagger}(0) U(1)]\boldsymbol{)}$. 
Note also that this dynamical phase coincides with the one in Eq. 
(14) of Ref. \cite{sjoqvist00} for the maximally mixed internal 
state $\frac{1}{D} \hat{1}$ in an interferometer.

The channel holonomy related to the family of unitary channels can
alternatively be calculated using the parallel transport gauge, as
developed in Sec. \ref{sec:smooth}. The parallel transport condition 
in Eq. (\ref{eq:smoothparallel}) reduces to the requirement that
$\Tr[\dot{\widetilde{U}}^{\dagger}(s)\widetilde{U} (s)]$ should be 
a real number for all $s$. However, from the unitarity of 
$\widetilde{U}(s)$ it follows that this number can only be purely 
imaginary. We thus find the parallel transport condition 
\begin{equation}
\Tr[\dot{\widetilde{U}}^{\dagger}(s)\widetilde{U}(s)] = 0.
\label{eq:ptunitary} 
\end{equation}
This condition is satisfied by the solution of 
the equation of motion 
$i\dot{\widetilde{U}}(s) = \{ H(s) - \frac{1}{D} \Tr[H(s)] 
\hat{1} \} \widetilde{U}(s)$. The unitaries $U(s)$ and 
$\widetilde{U}(s)$ are related by the gauge transformation 
\begin{eqnarray}
U(s) \rightarrow \widetilde{U}(s) = 
U(s) \exp \left[ \frac{i}{D} \int_0^s ds' \Tr H(s') .  
\right] 
\label{eq:relation}
\end{eqnarray} 
The holonomy in the parallel transport gauge  
takes the form 
\begin{eqnarray} 
\gamma_{\textrm{\textrm{ch}}} (C) = 
\Phi \boldsymbol{(} \Tr [\widetilde{U}^{\dagger} (0) 
\widetilde{U}(1) ] \boldsymbol{)} ,  
\label{eq:chptunitary}
\end{eqnarray}   
which can be seen by inserting Eq. (\ref{eq:relation}) into 
Eq. (\ref{eq:kinunitary}). 

Let us finally consider unitary channels for a single qubit (i.e.,
$D=2$). We claim that one-qubit holonomies only can take the values
$\pm 1$. To see this, we first note that U(2)=SU(2)$\times$U(1). The
U(1) parts of the sequence cannot contribute to the holonomy due to
their cyclic appearance on the right-hand side of
Eq. (\ref{eq:utwist}). The claim then follows from
$\Phi\boldsymbol{(}\Tr[\textrm{SU(2)SU(2)}]
\boldsymbol{)} = \Phi\boldsymbol{(}\Tr [\textrm{SU(2)}]\boldsymbol{)}=\pm 1$, 
since $\Tr [\textrm{SU(2)}]$ is a real number \cite{remark4}. 

\subsection{Holonomic channels}
Consider a smoothly parametrized decomposition $\mathcal{H}_q = 
\bigoplus_{k=1}^K \mathcal{H}_k(s')$ of a $D$-dimensional 
Hilbert space $\mathcal{H}_q$, for $s'\in [0,s]$. Assume 
$\dim \mathcal{H}_k(s')=D_k$ is constant for all $k$ on the 
interval $[0,s]$ and let $P_k(s')$ be the projection 
operator onto $\mathcal{H}_k(s')$. The quantities
\begin{eqnarray}
\Gamma_k (s)= \lim_{\delta s \rightarrow 0} P_k(s)
P_k(s-\delta s)\ldots P_k(0) 
\end{eqnarray}
can be expressed as \cite{kult06}
\begin{eqnarray}
\label{Gammaexp}
\Gamma_k (s) = \sum_{ij} [{\bf P}e^{\int_0^{s} 
\boldsymbol{A}_k(s')ds'}]_{ij} |a_i^{(k)} (s) \rangle 
\langle a_j^{(k)} (0)|,
\end{eqnarray}
where ${\bf P}$ denotes path ordering, the anti-Hermitian matrix 
$\boldsymbol{A}_k (s')$ has elements $[\boldsymbol{A}_k(s')]_{ij} = 
\langle \dot{a}_i^{(k)} (s')|a_j^{(k)} (s') \rangle$, and 
$\{|a_i^{(k)} (s')\rangle \}_{i=1}^{D_k}$ is an orthonormal 
basis of $\mathcal{H}_k(s')$. A holonomic channel is defined as 
\begin{eqnarray}
\mathcal{E}_{s}^{{\textrm{hol}}} (\rho) = 
\sum_k \Gamma_k (s)\rho \Gamma_k^\dagger (s).
\end{eqnarray}
The condition for trace preservation 
$\sum_k \Gamma_k^\dagger (s)\Gamma_k (s) = \hat{1}$ can be
shown to be satisfied by using that ${\bf P} e^{\int_0^{s} 
\boldsymbol{A}_k (s')ds'}$ in Eq. (\ref{Gammaexp}) 
is a unitary matrix.

Let us now examine the holonomy corresponding to a curve 
$C: [0,1] \ni s \mapsto \mathcal{E}_{s}^{{\textrm{hol}}}$ 
of holonomic channels. The parallel transport 
condition in Eq. (\ref{eq:smoothparallel}) is satisfied if 
the matrix with elements $\Tr[\dot{\Gamma}_{k}^{\dagger}(s)
\Gamma_{l}(s)]$ is Hermitian. By a direct use of Eq. (\ref{Gammaexp}) 
we find that $\Tr[\dot{\Gamma}_{k}^{\dagger}(s)\Gamma_{l}(s)] = 0$ 
for all $k,l$ and $s$. We can thus conclude that 
$\{\Gamma_{k}(s)\}_{k}$ is parallel transported. Consequently, 
the holonomy is given by
\begin{eqnarray}
\boldsymbol{U}_{\textrm{\textrm{ch}}} (C) = \Phi(\boldsymbol{T}_{0,1}) . 
\end{eqnarray}
Here, $\boldsymbol{T}_{0,1}$ is a matrix with elements 
\begin{eqnarray}
[\boldsymbol{T}_{0,1}]_{kl} & = & 
\Tr [ P_k(0)\Gamma_l (1) ] 
\nonumber \\ 
& = & \delta_{kl} \sum_{ij} [{\bf P} 
e^{\int_0^{1} \boldsymbol{A}_k (s)ds}]_{ij} 
\langle a_j^{(k)}(0)|a_i^{(k)} (1) \rangle 
\nonumber \\ 
 & = & \delta_{kl} \Tr 
[ \boldsymbol{U}_g (\mathcal{C}_k) ] ,  
\end{eqnarray}
where $\boldsymbol{U}_g (\mathcal{C}_k)$ is the holonomy of 
the path $\mathcal{C}_k$ in the Grassmann manifold $\mathcal{G} (D;D_k)$, 
i.e., the space of $D_k$-dimensional subspaces of an $D$-dimensional 
Hilbert space. Thus, the holonomy takes the form 
\begin{eqnarray} 
\boldsymbol{U}_{\textrm{\textrm{ch}}} (C) & = & 
\textrm{diag} \{ \Phi \boldsymbol{(}\Tr [\boldsymbol{U}_g 
(\mathcal{C}_1) ] \boldsymbol{)}, \Phi \boldsymbol{(} \Tr [ \boldsymbol{U}_g 
(\mathcal{C}_2) ] \boldsymbol{)},  
\nonumber \\ 
 & & \ldots , \Phi \boldsymbol{(} \Tr [ \boldsymbol{U}_g 
(\mathcal{C}_K)  ] \boldsymbol{)}\} .  
\end{eqnarray} 
Note that if $\mathcal{C}_k$ is a closed path, 
$[\boldsymbol{T}_{0,1}]_{kk}$ equals the Wilson loop 
$\Tr {\bf P} e^{\oint_{\mathcal{C}_k} \boldsymbol{A}_k (s)ds}$ 
\cite{zee88}.  

A possible way to implement the holonomic channels is to approximate 
them by sequences of projective measurements. We discretize the 
interval $[0,s]$ with step size $\delta s$. We first perform the 
measurement $\{ P_{k_0} (0) \}_{k_0}$ followed by $\{ P_{k_1} (\delta s) 
\}_{k_1}$ up to $\{ P_{k_N} (s) \}_{k_N}$, where $s=N\delta s$. 
Discarding the outcomes of these measurements the resulting 
operation on the input density operator $\rho$ reads 
\begin{eqnarray}
\label{sequence}
\rho & \rightarrow & \sum_{k_0,\ldots,k_N} P_{k_N}(N\delta s) 
\ldots P_{k_0}(0) \rho P_{k_0}(0) \ldots P_{k_N} (N\delta s) 
\nonumber \\ 
 & = & \sum_k P_k (N\delta s) 
\ldots P_k (0) \rho P_k (0) \ldots P_k (N\delta s) 
\nonumber \\ 
 & & + \mathcal{R} (\rho) \equiv \mathcal{M} (\rho) ,   
\end{eqnarray} 
which is a channel. Here, 
\begin{eqnarray} 
\mathcal{R} (\rho) & = & 
\sum_{k_0,\ldots,k_N \in \mathcal{K}} P_{k_N}(N\delta s) 
\ldots P_{k_0}(0) \rho 
\nonumber \\ 
 & & \times P_{k_0}(0) \ldots P_{k_N} (N\delta s) , 
\end{eqnarray}
where $\mathcal{K}$ is the set of values of $k_0,\ldots,k_N$ 
excluding those where $k_0 = \ldots = k_N$. Note that 
both $\sum_k P_k (N\delta s) 
\ldots P_k (0) \rho P_k (0) \ldots P_k (N\delta s)$ 
and $\mathcal{R} (\rho)$ correspond to completely positive 
maps. In the $\delta s \rightarrow 0$ limit we have  
\begin{eqnarray}
\lim_{\delta s \rightarrow 0} \mathcal{M} (\rho) = 
\mathcal{E}_{s'}^{{\textrm{hol}}} (\rho) + 
\lim_{\delta s \rightarrow 0} \mathcal{R} (\rho) . 
\end{eqnarray}
Now, $\mathcal{M}$ and $\mathcal{E}_{s'}^{{\textrm{hol}}}$ are trace 
preserving from which it follows that $\mathcal{R}$ must vanish in 
the $\delta s \rightarrow 0$ limit. Thus, we have 
\begin{eqnarray}
\lim_{\delta s \rightarrow 0} \mathcal{M} (\rho) = 
\mathcal{E}_{s'}^{{\textrm{hol}}} (\rho) , 
\end{eqnarray}
which concludes our demonstration that the holonomic channels 
can be approximated by sequences of projective measurements. 

\section{Physical realization}
\label{sec:realization}  
Given a physical system and a sequence of operations
$\mathcal{E}_{1},\ldots,\mathcal{E}_{N}$ acting on this system, one
might ask: what is the physical significance of the channel holonomy,
and how should it be measured? Strictly speaking, these two questions
have no well defined answer, given how we have constructed the channel
holonomy. We have defined it as a change in Kraus representation, but
the Kraus representation as such has no direct physical
significance. Hence, there is no immediate way to attach measurable
quantities to the channel holonomy. In other words, up to this point
the channel holonomy has been a mathematical construction, rather than
corresponding directly to a physical object or operation. However, we
shall here obtain such a correspondence within the context of
interferometry. Another related question is, what is the object that
the channel holonomy $\boldsymbol{U}_{\textrm{ch}}$ represents?  As
$\boldsymbol{U}_{\textrm{ch}}$ is a gauge-covariant unitary matrix, it
seems intuitively reasonable that it should be a matrix representation
of a gauge-invariant object. Our interferometric construction will
provide precisely such an object.

\subsection{Yet another parallel transport}
\label{sec:yet}
Every channel $\mathcal{E}$ on a Hilbert space $\mathcal{H}_q$ with
dimension $D$ can be obtained via a joint unitary evolution on the
system and an ancillary system with Hilbert space $\mathcal{H}_a$,
according to 
\begin{equation}
\label{repr}
\mathcal{E}(\rho) =  
\Tr_a [\mathcal{U}(\rho\otimes|a\rangle\langle a|) 
\mathcal{U}^{\dagger}],
\end{equation} 
where $\mathcal{U}$ is a unitary operator on $\mathcal{H}_q \otimes 
\mathcal{H}_a$ and $|a\rangle\in\mathcal{H}_a$ is normalized. If $U$ 
is an arbitrary unitary operator on $\mathcal{H}_a$, then both
$\mathcal{U}$ and $(\hat{1}\otimes U)\mathcal{U}$ give rise to the
same channel $\mathcal{E}$ \cite{note}. If we regard the set of
unitary operators $\mathcal{U}$ representing $\mathcal{E}$ as a fiber,
the transformation $\mathcal{U}\rightarrow (\hat{1}\otimes
U)\mathcal{U}$ can be seen as a gauge transformation resulting in
a motion along the fiber.

Consider now the sequence of channels $\mathcal{E}_{1},\ldots, 
\mathcal{E}_{N}$, all with Kraus number $K$.  We assume 
an ancillary space $\mathcal{H}_{a}$ of dimension $K$, and consider 
a sequence of unitary operators $\widetilde{\mathcal{U}}_{1},\ldots,
\widetilde{\mathcal{U}}_{N}$ on $\mathcal{H}_{q}\otimes\mathcal{H}_{a}$, 
where $\widetilde{\mathcal{U}}_{n}$ represents $\mathcal{E}_{n}$ via
Eq. (\ref{repr}). We regard the sequence of unitaries as parallel
transported if
\begin{equation}
\label{paratransp}
\Tr_{q}[\widetilde{\mathcal{U}}_{n}(\hat{1}\otimes 
|a\rangle\langle a|)\widetilde{\mathcal{U}}_{n+1}^{\dagger}] >0. 
\end{equation}
Given an arbitrary sequence of unitaries
$\mathcal{U}_{1},\ldots, \mathcal{U}_{N}$ we can make
it into a parallel transported sequence $\widetilde{\mathcal{U}}_{n}
= (\hat{1}\otimes U_{n})\mathcal{U}_{n}$ by choosing unitary 
operators $U_{1},\ldots, U_{N}$ such that
\begin{eqnarray}
\label{procedure}
 & & \Tr_{q}[\widetilde{\mathcal{U}}_{n}(\hat{1}\otimes 
|a\rangle\langle a|)\widetilde{\mathcal{U}}_{n+1}^{\dagger}] 
\nonumber \\ 
 & & = U_{n}\Tr_{q}[\mathcal{U}_{n}(\hat{1}\otimes |a\rangle\langle a|) 
\mathcal{U}_{n+1}^{\dagger}]U_{n+1}^{\dagger}>0.
\end{eqnarray}
This requires that $\Tr_{q}[\mathcal{U}_{n}(\hat{1}\otimes
|a\rangle\langle a|)\mathcal{U}_{n+1}^{\dagger}]$ is of rank $K$. 
It then follows that
\begin{eqnarray}
\label{UNp1}
U_{N} & = & U_{1}\Phi\boldsymbol{(}\Tr_{q}[\mathcal{U}_{1}
(\hat{1}\otimes|a\rangle\langle
a|)\mathcal{U}_{2}^{\dagger}]\boldsymbol{)}\cdots
\nonumber\\
 & & \cdots \Phi\boldsymbol{(}\Tr_{q}[\mathcal{U}_{N-1}(\hat{1}\otimes
|a\rangle\langle a|)\mathcal{U}_{N}^{\dagger}]\boldsymbol{)}
\end{eqnarray}
and $\widetilde{\mathcal{U}}_{N} = 
(\hat{1} \otimes U_{N}) \mathcal{U}_{N}$.

As may be seen, the above construction strongly resembles the channel
holonomy in Sec. \ref{sec:chaho}, as well as the Uhlmann holonomy in
Sec. \ref{sec:jamiolkowski}. We shall see that this is not a mere
coincidence, but that the above construction enables us to obtain
the channel holonomy within the context of interferometry.  With this
purpose in mind we review some of the concepts and tools that are useful
for the analysis of quantum operations in single-particle interferometry.
A more thorough account of these theoretical 
tools can be found in Refs. \cite{aberg04a,aberg04b}. See also 
Refs. \cite{oi03,oi06,aberg07} for related material.  

\subsection{Operations in interferometry}
Suppose that we have a single particle with some internal degrees of
freedom (e.g., spin or polarization) with Hilbert space
$\mathcal{H}_{q}$, and that the particle can propagate along two
separated paths (e.g., the two paths in a Mach-Zehnder
interferometer). These two paths correspond to two orthonormal vectors
$|0\rangle$ and $|1\rangle$ spanning the ``spatial'' Hilbert space
$\mathcal{H}_p = \Sp\{|0\rangle, |1\rangle\}$. Thus, the total Hilbert
space of the particle is $\mathcal{H}_p \otimes
\mathcal{H}_q$. 

Now, suppose that we have two operations $\Lambda_{0}$ and
$\Lambda_{1}$ acting on the internal degrees of freedom of the
particle. Let $\Lambda_{0}$ operate on the particle if it passes path
$0$, and let $\Lambda_{1}$ operate on the particle if it passes path
$1$. The question is, what channels $\Lambda$ acting on elements of
$\mathcal{L} (\mathcal{H}_p\otimes \mathcal{H}_q)$ would be compatible
with the channels $\Lambda_{0}$ and $\Lambda_{1}$ acting in each path? 
(To be more precise, we require $\Lambda$ to be a channel such that
$\Lambda(|0\rangle\langle0|\otimes\rho) = |0\rangle\langle
0|\otimes\Lambda_{0}(\rho)$ and $
\Lambda(|1\rangle\langle1|\otimes\rho) = |1\rangle\langle
1|\otimes\Lambda_{1}(\rho)$, for all density operators $\rho$ on
$\mathcal{H}_q$.) The answer is as follows. Let $\{V_{n}\}_{n}$ and
$\{W_{m}\}_{m}$ be linearly independent Kraus representations of the 
channels $\Lambda_{0}$ and $\Lambda_{1}$, respectively. Then 
\cite{aberg04a,aberg04b}
\begin{eqnarray}
\Lambda(\sigma) & = & 
|0\rangle\langle 0|\otimes\Lambda_{0}(\langle 0|\sigma|0\rangle) + 
|1\rangle\langle 1|\otimes\Lambda_{1}(\langle 1|\sigma|1\rangle) 
\nonumber\\
 & & + |0\rangle\langle 1|\otimes\sum_{nm}\boldsymbol{C}_{n,m} 
V_{n}\langle 0|\sigma|1\rangle W_{m}^{\dagger} 
\nonumber\\
 & & + |1\rangle\langle 0|\otimes\sum_{nm}\boldsymbol{C}_{n,m}^{*} 
W_{m}\langle 1|\sigma|0\rangle V_{n}^{\dagger}
\end{eqnarray}
where the matrix $\boldsymbol{C}$ satisfies 
$\boldsymbol{C}\boldsymbol{C}^{\dagger} \leq \boldsymbol{I}$ and 
$\sigma \in \mathcal{L} (\mathcal{H}_p \otimes \mathcal{H}_q)$. 
We refer to $\Lambda$ as a ``gluing'' of the two channels 
$\Lambda_{0}$ and $\Lambda_{1}$, and to the matrix $\boldsymbol{C}$ 
as the ``gluing matrix''.

All gluings can be obtained using a shared ancillary system between
the two paths \cite{aberg04a,aberg04b}. Let $\mathcal{V}^{(0)}$ and
$\mathcal{V}^{(1)}$ be unitary operators representing the channels
$\Lambda_{0}$ and $\Lambda_{1}$ via Eq. (\ref{repr}), where we assume
the same ancilla.  On the combined system of the two paths, the
system, and the ancilla, we can construct the unitary operator
\begin{equation}
\label{Utot}
U_{\textrm{tot}} = 
|0\rangle\langle 0|\otimes \mathcal{V}^{(0)} + 
|1\rangle\langle 1|\otimes\mathcal{V}^{(1)} .  
\end{equation}
It turns out that 
\begin{equation}
\Lambda(\sigma) = 
\Tr_{a}(U_{\textrm{tot}} \sigma \otimes 
|a\rangle\langle a|U^{\dagger}_{\textrm{tot}})
\end{equation}
is a gluing of the channels $\Lambda_{0}$ and $\Lambda_{1}$, and
moreover, that every gluing can be obtained by the appropriate choices
of $\mathcal{V}^{(0)}$ and $\mathcal{V}^{(1)}$ \cite{aberg04a,
aberg04b}. Hence, although various choices of $\mathcal{V}^{(0)}$ and
$\mathcal{V}^{(1)}$ represent the same channels $\Lambda_{0}$ and
$\Lambda_{1}$, respectively, these choices may nevertheless 
result in different gluings of $\Lambda_{0}$ and $\Lambda_{1}$.

\subsection{Physical realization of the channel holonomy}
\label{sec:interreal} 
\subsubsection{Interferometric parallel transport procedure}
\label{sec:measureparall} 
We now consider the Mach-Zehnder setup. Let the particle start in
path $0$ and internal state $\rho\otimes|a\rangle\langle a|$, after
which we apply a 50-50 beam splitter on the spatial degrees of freedom,
acting as a Hadamard gate $\mathbb{H}$ on the spatial degrees of freedom
regarded as a qubit. Thereafter, we apply $U_{\textrm{tot}}$ in 
Eq. (\ref{Utot}) on the total system followed by a variable unitary 
operator $V^{(0)}$ on the ancillary Hilbert space in path $0$ and 
a variable unitary operator $V^{(1)}$ in path $1$, i.e., the unitary 
operator
\begin{equation}
F = |0\rangle\langle 0|\otimes \hat{1}\otimes V^{(0)} + 
|1\rangle\langle 1|\otimes \hat{1}\otimes V^{(1)}.
\end{equation}
Finally, we apply a second beam splitter and calculate the expectation
value of the projector $|0\rangle\langle 0|\otimes \hat{1}\otimes
\hat{1}$, i.e., the probability $p$ to find the particle in path 0, 
yielding 
\begin{eqnarray}
p & = & \Tr[(|0\rangle\langle 0|\otimes\hat{1}\otimes\hat{1}) 
\mathbb{H}FU_{\textrm{tot}}
\nonumber \\
 & & \times \mathbb{H}(|0\rangle\langle 0|\otimes\rho \otimes 
|a\rangle\langle a|) \mathbb{H}^{\dagger} U_{\textrm{tot}}^{\dagger} 
F^{\dagger}\mathbb{H}^{\dagger}] 
\nonumber\\
 & = & \frac{1}{2} + \frac{1}{2}\Real\Tr\{V^{(0)} 
\Tr_{q}[\mathcal{V}^{(0)}(\rho\otimes|a\rangle\langle a|)
{\mathcal{V}^{(1)}}^{\dagger}]{V^{(1)}}^{\dagger} \} . 
\nonumber \\ 
\end{eqnarray}
If we regard the unitary operator $V^{(0)}$ as fixed, then we find
that the maximum probability $p$ is obtained when $V^{(1)} =
V^{(0)}\Phi\boldsymbol{(}\Tr_{q}[\mathcal{V}^{(0)}(\rho\otimes 
|a\rangle\langle a|){\mathcal{V}^{(1)}}^{\dagger}]\boldsymbol{)}$.  
In the special case where $\rho = \frac{1}{D} \hat{1}$, i.e., the 
initial internal state is maximally mixed, we thus find that 
\begin{equation}
V^{(1)} = V^{(0)} \Phi\boldsymbol{(} 
\Tr_{q}[\mathcal{V}^{(0)}(\hat{1} \otimes |a\rangle\langle a|) 
{\mathcal{V}^{(1)}}^{\dagger}]\boldsymbol{)} 
\end{equation} 
maximizes the detection probability $p$. 

Comparing with Eq. (\ref{procedure}) we find that the parallel
transport procedure can be implemented using this interferometric
approach.  We have a sequence of unitaries $\mathcal{U}_{1},\ldots,
\mathcal{U}_{N}$.  We let $\mathcal{V}^{(0)} = \mathcal{U}_{1}$ and
$\mathcal{V}^{(1)} = \mathcal{U}_{2}$, and choose the initial 
unitary operator $V^{(0)} = U_{1}$ arbitrarily. Next, we let the input
internal state to the interferometer be the maximally mixed state
$\rho = \frac{1}{D} \hat{1}$ and vary $V^{(1)}$ until maximal detection
probability is obtained.  The maximum is achieved when $V^{(1)} =
U_{2}$. In the next step we repeat the process but let
$\mathcal{V}^{(0)} = \mathcal{U}_{2}$ and $\mathcal{V}^{(1)} =
\mathcal{U}_{3}$ and we let $V^{(0)} = U_{2}$, i.e., the unitary
operator we obtained in the first step. We then vary $V^{(1)}$ until
we obtain maximal detection probability. We repeat this procedure up
to $\mathcal{U}_{N}$. 

One may note that the parallel transport procedure just described is 
purely operational, in the sense that it is achieved as a result 
of optimizing detection probabilities. This is analogous to the 
approach to the Uhlmann holonomy as developed in Ref. \cite{aberg07}.

\subsubsection{Realizing the channel holonomy 
$\boldsymbol{U}_{\textrm{ch}}$ as a gluing between the 
end point channels}
\label{sec:realchaho} 
After the $N$th step of the parallel transport procedure, we can 
construct the channel holonomy as a measurable object in the
interferometer. We let $\mathcal{V}^{(0)} =
\mathcal{U}_{N}$,  $V^{(0)} =
U_{N}$, $\mathcal{V}^{(1)} = \mathcal{U}_{1}$, and $V^{(1)} = U_{1}$.
If we discard the ancillary system the resulting channel on 
$\mathcal{L}(\mathcal{H}_p \otimes \mathcal{H}_q)$ is
\begin{eqnarray}
\label{finaldelsteg}
\Lambda_{\textrm{final}}(\sigma) = & 
|0\rangle\langle 0|\otimes \mathcal{E}_{N} 
(\langle 0|\sigma |0\rangle) + 
|1\rangle\langle 1|\otimes \mathcal{E}_{1} 
(\langle 1|\sigma |1\rangle) 
\nonumber\\
 & + |0\rangle\langle 1|\otimes \mathcal{G}(\sigma) + 
|1\rangle\langle 0|\otimes \mathcal{G}^{\dagger}(\sigma),
\end{eqnarray}
where 
\begin{equation}
\label{Gdef}
\mathcal{G}(\sigma) = \Tr_{a}[(\hat{1}\otimes U_{N})\mathcal{U}_{N} 
(\langle 0|\sigma |1\rangle\otimes |a\rangle\langle a|) 
\mathcal{U}_{1}^{\dagger}(\hat{1}\otimes U_{1}^{\dagger})] . 
\end{equation}
We shall now see that the channel $\Lambda_{\textrm{final}}$ is
directly related to the channel holonomy.  

Consider a unitary operator $\mathcal{U}_{n}$ representing
$\mathcal{E}_{n}$ via Eq. (\ref{repr}). Choose an arbitrary but fixed
orthonormal basis $\{|a_{k}\rangle\}_{k=1}^{K}$ of $\mathcal{H}_{a}$
for the evaluation of $\Tr_{a}$. For such a choice, the operators 
$E_{k}^{n} = \langle a_{k}|\mathcal{U}_{n}|a\rangle$ form a 
linearly independent Kraus representation of the channel
$\mathcal{E}_{n}$. We find
\begin{equation}
\label{Gomsk}
\mathcal{G}(\sigma) =  
\sum_{k,k'}\langle a_{k}|U_{1}^{\dagger} 
U_{N}|a_{k'}\rangle E_{k'}^{N}\langle 0|\sigma |1\rangle 
{E_{k}^{1}}^{\dagger}.
\end{equation}
One can furthermore show 
\begin{equation}
\label{overs}
\langle a_{k}|\Tr_{q}[\mathcal{U}_{n}(\hat{1}\otimes|a\rangle\langle a|)
\mathcal{U}^{\dagger}_{n+1}]|a_{k'}\rangle = [\boldsymbol{T}_{n+1,n}]_{k'k}.
\end{equation}
(Note the reordering of $k$ and $k'$ between the left- and right-hand
side.) As can be seen from the above equation, 
$\Tr_{q}[\mathcal{U}_{n}(\hat{1}\otimes|a\rangle\langle
a|)\mathcal{U}^{\dagger}_{n+1}]$ is of rank $K$ if and only if
$\boldsymbol{T}_{n+1,n}$ is of rank $K$. Equation (\ref{overs})
implies 
\begin{eqnarray}
 & \langle a_{k}| \Phi\boldsymbol{(} \Tr_{q} [\mathcal{U}_{n} 
(\hat{1}\otimes|a\rangle\langle a|)\mathcal{U}^{\dagger}_{n+1}] 
\boldsymbol{)}|a_{k'}\rangle 
\nonumber \\ 
 & = [\Phi(\boldsymbol{T}_{n+1,n})]_{k'k} ,  
\end{eqnarray}
which can be combined with Eq. (\ref{UNp1}) to give   
\begin{eqnarray}
\langle a_{k}|U_{1}^{\dagger}U_{N}|a_{k'}\rangle & = & 
[\Phi(\boldsymbol{T}_{N,N-1})\ldots\Phi(\boldsymbol{T}_{2,1})]_{k'k} 
\nonumber \\ 
 & \equiv & \boldsymbol{Z}_{k'k}. 
\end{eqnarray}
It follows that 
\begin{eqnarray}
\label{nastanklar}
\Lambda_{\textrm{final}}(\sigma) & = & 
|0\rangle\langle 0|\otimes \mathcal{E}_{N}(\langle 0|\sigma
|0\rangle) + |1\rangle\langle 1|\otimes \mathcal{E}_{1}(\langle
1|\sigma |1\rangle) 
\nonumber\\ 
 & & + |0\rangle\langle 1|\otimes
\sum_{k,k'=1}^{K}\boldsymbol{Z}_{k'k}
E_{k'}^{N}\langle 0|\sigma |1\rangle {E_{k}^{1}}^{\dagger} 
\nonumber\\
 & & + |1\rangle\langle
0|\otimes\sum_{k,k'=1}^{K} E_{k'}^{1} \langle 1|\sigma |0\rangle 
{E_{k}^{N}}^{\dagger} \boldsymbol{Z}^{\dagger}_{kk'} , 
\end{eqnarray}
where the two Kraus representations $\{E_{k}^{1}\}_{k}$ and
$\{E_{k}^{N}\}_{k}$ are free and independent of each other. We may
therefore choose a Kraus representation $\{\overline{E}_{l}^{N}\}_{l}$
of the channel $\mathcal{E}_{N}$, such that the overlap matrix with
elements $\Tr({\overline{E}_{l}^{N}}^{\dagger}E^{1}_{k})$ is positive
definite, i.e., when the Kraus representation of $\mathcal{E}_{N}$ is
parallel with the initial Kraus representation, which we know is
possible if the overlap matrix $[\boldsymbol{T}_{1,N}]_{lk} =
\Tr({E_{l}^{1}}^{\dagger}E^{N}_{k})$ is invertible. We thus substitute
$E^{N}_{k'} =
\sum_{k}\overline{E}^{N}_{k}[\Phi(\boldsymbol{T}_{1,N})]_{kk'}$ into
Eq. (\ref{nastanklar}), which yields
\begin{eqnarray}
\label{eq:finalfinal}
\Lambda_{\textrm{final}}(\sigma) & = & 
|0\rangle\langle 0|\otimes \mathcal{E}_{N}(\langle 0|\sigma
|0\rangle) + |1\rangle\langle 1|\otimes \mathcal{E}_{1}(\langle
1|\sigma |1\rangle) 
\nonumber \\ 
 & & + |0\rangle\langle 1|\otimes
\sum_{k,k'=1}^{K}[\boldsymbol{U}_{\textrm{ch}}]_{k'k}
\overline{E}_{k'}^{N}\langle 0|\sigma |1\rangle {E_{k}^{1}}^{\dagger} 
\nonumber\\
 & & + |1\rangle\langle0| \otimes \sum_{k,k'=1}^{K} 
 E_{k'}^{1}\langle 1|\sigma |0\rangle 
{\overline{E}_{k}^{N}}^{\dagger} 
[\boldsymbol{U}_{\textrm{ch}}^{\dagger}]_{kk'} . 
\nonumber\\
\end{eqnarray}
Thus, the channel holonomy
$\boldsymbol{U}_{\textrm{ch}}$ is nothing but the gluing matrix
describing the gluing of the two end point channels $\mathcal{E}_{1}$
and $\mathcal{E}_{N}$, with respect to the choice of 
parallel Kraus representations of these two channels. It is to be
noted that a gauge transformation $E^{1}_{k}\rightarrow
\sum_{l}E^{1}_{l}\boldsymbol{V}_{lk}$ imply the same transformation
$\overline{E}^{N}_{k}\rightarrow \sum_{l}\overline{E}^{N}_{l} 
\boldsymbol{V}_{lk}$ due to the assumption of parallelity 
between the end points. One may also note that the gauge covariance of
$\boldsymbol{U}_{\textrm{ch}}$, as described in Eq. (\ref{covariant})
is necessary for $\Lambda_{\textrm{final}}$ to be gauge
invariant. Another way to put this is to say that
$\boldsymbol{U}_{\textrm{ch}}$ in some sense is the matrix
representation of the gluing $\Lambda_{\textrm{final}}$, and as a
matrix representation of a gauge-invariant object we expect
$\boldsymbol{U}_{\textrm{ch}}$ to be gauge covariant.  We also point
out that in the special case where the two end point channels
coincide, i.e., $\mathcal{E}_N = \mathcal{E}_1$, then 
$\overline{E}^{N}_{k} = E^{1}_{k}$.

The above construction of the channel holonomy as a matrix
representation with respect to two parallel Kraus
representations is analogous to the construction of the 
open-path holonomy in Ref. \cite{kult06}. This is perhaps best 
seen in Eq. (12) in Ref. \cite{kult06}, where the open-path 
holonomy $\boldsymbol{U}_{g}$ can be expressed as $\Gamma =
\sum_{kl}[\boldsymbol{U}_{g}]_{kl}|\overline{a}_{k}(1)\rangle\langle
a_{l}(0)|$. Here $\{|\overline{a}_{k}(1)\rangle\}_{k}$ is a frame
which is parallel to the frame $\{|a_{k}(0)\rangle\}_{k}$ in
a sense that resembles the parallelity of Kraus
representations. (Compare, e.g., Eq. (\ref{maxparall}) of the 
present paper with Eq. (8) in Ref. \cite{kult06}.)

We finally note that the gluing matrix is in principle possible to
measure using interferometric setups \cite{aberg04b}.  Hence, the
channel holonomy resulting from the parallel transport procedure in
Sec. \ref{sec:measureparall} is a measurable object. 

\section{The case of smooth parametrization}
\label{sec:realizationsmooth}
In Sec. \ref{sec:smooth} we considered the smooth version of the
iterative parallel transport in Sec. \ref{sec:chaho}. Here we perform the
analogous transition for the ancillary construction in Sec. \ref{sec:yet},
finding conditions for parallel transport.

Consider a smooth family of unitary operators
$\widetilde{\mathcal{U}}(s)$ acting on the combined system and ancilla
$\mathcal{H}_{q}\otimes\mathcal{H}_{a}$, thus generating a smooth
family of channels
\begin{equation}
\label{kanaldef}
\mathcal{E}_s (\rho) = 
\Tr_{a}[\widetilde{\mathcal{U}} (s) \rho \otimes |a\rangle\langle a| 
\widetilde{\mathcal{U}}^{\dagger}(s)].
\end{equation}
When can we say that the family $\widetilde{\mathcal{U}}(s)$ is
parallel transported? By recalling Eq. (\ref{paratransp}) it appears
reasonable to require that
\begin{equation}
\Tr_{q}[\widetilde{\mathcal{U}}(s) 
(\hat{1}\otimes |a\rangle\langle a|)
\widetilde{\mathcal{U}}^{\dagger}(s+ \delta s)]  >0
\end{equation}
to the first order in $\delta s$ in the limit of small $\delta s$. 
We find
\begin{equation}
\Tr_{q}[\widetilde{\mathcal{U}}(s)(\hat{1}\otimes 
|a\rangle\langle a|)\widetilde{\mathcal{U}}^{\dagger}(s+ \delta s)] 
=  \mathcal{Q}(s) + \delta s \mathcal{R}(s),
\end{equation}
where
\begin{eqnarray}
\mathcal{Q}(s) = & \Tr_{q}[\widetilde{\mathcal{U}}(s) 
(\hat{1}\otimes |a\rangle\langle a|)
\widetilde{\mathcal{U}}^{\dagger}(s)], \\
\mathcal{R}(s) = & \Tr_{q}[\widetilde{\mathcal{U}}(s)
(\hat{1}\otimes |a\rangle\langle a|)
\dot{\widetilde{\mathcal{U}}}^{\dagger}(s)].
\end{eqnarray}
Note that $\mathcal{Q}(s) \geq 0$. If we further assume that 
\begin{equation}
\label{conditionsmooth}
\mathcal{Q}(s)>0,
\end{equation}
then we find that the parallel transport condition is satisfied 
when $\mathcal{R}(s)$ is Hermitian, i.e.,
\begin{eqnarray}
\label{condiparallel}
 & & \Tr_{q}[\widetilde{\mathcal{U}}(s)(\hat{1}\otimes |a\rangle\langle a|)
\dot{\widetilde{\mathcal{U}}}^{\dagger} (s)]
\nonumber \\ 
 & & = \Tr_{q}[\dot{\widetilde{\mathcal{U}}}(s)
(\hat{1}\otimes |a\rangle\langle a|)
\widetilde{\mathcal{U}}^{\dagger}(s)] . 
\end{eqnarray}

Suppose we have a smooth family of unitaries $\mathcal{U}(s)$ that is
not parallel transported. The question is, under what conditions we
can make it parallel transported by multiplying with a unitary on the
ancillary space
\begin{equation}
\label{ancillgaugetransf}
\widetilde{\mathcal{U}}(s) = [\hat{1} \otimes U(s)]\mathcal{U}(s),
\end{equation}
i.e., a gauge transformation that leaves the path of channels
$\mathcal{E}_s$ in Eq. (\ref{kanaldef}) unchanged? Here we elucidate when
it is possible to find a time-dependent Hamiltonian $H(s)$ on
$\mathcal{H}_{a}$ that generates $U(s)$ via a Schr\"odinger
equation. If we substitute Eq. (\ref{ancillgaugetransf}) into
Eq. (\ref{condiparallel}) we can rewrite the result as
\begin{equation}
\label{krav}
\begin{split}
\mathcal{R}(s)-\mathcal{R}^{\dagger}(s) =  
\mathcal{Q}(s) U^{\dagger}(s) \dot{U}(s) + 
U^{\dagger}(s)\dot{U}(s)\mathcal{Q}(s). 
\end{split}
\end{equation}
If we now consider an anti-Hermitian operator $\mathcal{A}(s)$
generating $U(s)$ via $\dot{U}(s) = U(s)\mathcal{A}(s)$, and
substitute into Eq. (\ref{krav}) we find
$\mathcal{R}-\mathcal{R}^{\dagger} = \mathcal{A}\mathcal{Q}
+\mathcal{Q}\mathcal{A}$.  Similarly as in Sec. \ref{sec:smooth} we
can use Theorem VII.2.3 in \cite{bhatia97} to conclude that there 
exists a unique anti-Hermitian operator $\mathcal{A}(s)$ solving this
equation. Hence, we can conclude that under the assumption
$\mathcal{Q}(s)>0$, there exists a unitary family $U(s)$ creating a
parallel transported family $\widetilde{\mathcal{U}}(s)$ via
Eq. (\ref{ancillgaugetransf}).

It is to be noted that we have described a kind of two-step 
procedure. First, the system and ancilla evolve jointly according to
$\mathcal{U}(s)$, which is not parallel transported. In the second
step, we modify the state by unitarily evolve the ancilla according to
$U(s)$. Hence, we have so far not obtained a joint Hamiltonian on the
system and ancilla that generates the parallel transported family
$\widetilde{\mathcal{U}}(s)$. However, the latter can in principle be
obtained since the Hamiltonian $\widetilde{H}(s) =
i\dot{\widetilde{\mathcal{U}}}\widetilde{\mathcal{U}}$ generates
$\widetilde{\mathcal{U}}$ via the Schr\"odinger equation. Hence, the
parallel transported evolution can in principle be tailored through a
combined evolution on the system and ancilla. Finally, we note 
that the parallel transport is not obtained operationally
in this smooth case, in the sense of the discrete case in
Sec. \ref{sec:interreal}. Whether such a ``interferometric parallel
transport procedure'' is possible in the smooth case we leave as an
open question.

\section{The $4\pi$ experiments}
\label{sec:4pi}
Bernstein \cite{bernstein67} and Aharonov and Susskind
\cite{aharonov67} pointed out the possibility to observe the $4\pi$
spinor symmetry of a spin-$\frac{1}{2}$ particle. The essence of their
argument was that this symmetry may show up in the relative phase, 
say between two paths in an interferometer, one in which the spinor is
rotated and one in which it is kept fixed. Subsequent neutron
interferometer experiments \cite{rauch75,werner75} were carried 
out to confirm this prediction.

These experiments used unpolarized neutrons that were sent through 
a two-beam interferometer, in which one beam was
exposed to a time-independent magnetic field ${\bf B}$ in the
$z$-direction, i.e., ${\bf B} = B\hat{\bf z}$. In the weak-field
limit, one can show that this corresponds to the family of unitary 
operations $U(s) = e^{-i\frac{s}{2} \sigma_z}, \ s\in[0,\varphi]$, 
where $\varphi \propto B$. A $4\pi$ periodic $\varphi$ oscillation 
in the intensity in one of the output beams was observed by varying 
$B$.

Let us consider the standard interpretation of these experiments. The
spinor part of unpolarized neutrons is described by the density
operator $\frac{1}{2} \hat{1}$, which can be decomposed into an
equal-weight mixture of any pair of orthogonal pure spin-$\frac{1}{2}$
states. Let the orthogonal vectors $\ket{\psi}$ and
$\ket{\psi^{\perp}}$ represent such a choice of states. These vectors
evolve into $U(\varphi) \ket{\psi}$ and $U(\varphi)
\ket{\psi^{\perp}}$, respectively, under the action of the magnetic
field. It follows that for $\varphi = n2\pi$, we obtain $\ket{\psi}
\rightarrow (-1)^n
\ket{\psi}$ and $\ket{\psi^{\perp}} \rightarrow (-1)^n
\ket{\psi^{\perp}}$. Thus, irrespective how we choose $\ket{\psi}$ and
$\ket{\psi^{\perp}}$, they have the desired $4\pi$ periodicity needed
to explain the experiments.

Here, we put forward another interpretation of these experiments,
based on the channel holonomy. We demonstrate that the
$4\pi$ periodic oscillations seen in the experiments can be
interpreted as a $4\pi$ periodicity of the channel holonomy in this
case. We also show that the channel holonomy itself can be measured, 
by a slight modification of the setup in Refs. \cite{rauch75,werner75}.

Let us first compute the holonomy of the path $C$ associated with the
family of unitary channels $\mathcal{E}_s^{{\textrm{u}}}$ represented 
by the unitary operators $U(s) = e^{-i\frac{s}{2} \sigma_z}, \ s\in
[0,\varphi]$. We obtain
\begin{eqnarray}
\Tr[\dot{U}^{\dagger} (s) U(s)] = 
-\frac{i}{2} \Tr [\sigma_z] = 0, 
\end{eqnarray}
i.e., $U(s)$ satisfies the parallel transport condition 
in Eq. (\ref{eq:ptunitary}). Thus, we may use Eq. 
(\ref{eq:chptunitary}) to deduce that 
\begin{eqnarray}
\gamma_{{\textrm{ch}}} [C] = 
\Phi\boldsymbol{(}\Tr [U(\varphi)] \boldsymbol{)},
\label{eq:4piholonomy}
\end{eqnarray} 
since $U(0)=\hat{1}$. Explicitly, $\Phi\boldsymbol{(}\Tr 
[U(\varphi)] \boldsymbol{)} = \Phi(\cos \frac{\varphi}{2})$, 
which is $4\pi$ periodic in $\varphi$. 

Next, we analyze the experimental setup in Refs.
\cite{rauch75,werner75} from the channel holonomy perspective. 
Let $\ket{0},\ket{1}$ represent the two beam directions. These vectors
constitute a basis for the spatial Hilbert space
$\mathcal{H}_{p}$. The experiment uses the standard interferometric
sequence with an initial beam splitter, followed by an operation in
the two paths, and finally a second beam splitter and a measurement.
For the moment we focus only on the operation occurring in between the
two beam splitters, but return below to the interferometer as a
whole. Let $\sigma$ be some arbitrary total state on
$\mathcal{H}_{p}\otimes\mathcal{H}_{q}$. Apply to the $0$ beam the
parallel transporting unitary family $U(s)$, $s\in [0,\varphi]$, that
represents the family of unitary channels along $C$. This results in
the map
\begin{eqnarray} 
 \Lambda^{{\textrm{u}}} (\sigma)  
 & = & \ket{0} \bra{0} \otimes \mathcal{E}_{\varphi}^{{\textrm{u}}} 
(\bra{0}\sigma\ket{0}) + \ket{1} \bra{1} \otimes 
\mathcal{E}_0^{{\textrm{u}}} (\bra{1}\sigma\ket{1}) 
\nonumber \\ 
 & & + \ket{0} \bra{1} \otimes   \gamma_{{\textrm{ch}}} [C] 
\overline{U} (\varphi) \bra{0} \sigma \ket{1} 
\nonumber \\
 & & + \ket{1} \bra{0} \otimes  
\bra{1} \sigma \ket{0}\overline{U}^{\dagger} (\varphi) 
\gamma_{{\textrm{ch}}} [C]^{*} ,   
\end{eqnarray}
where we have used Eq.~(\ref{eq:4piholonomy}), and where 
$\overline{U} (\varphi) = \Phi \boldsymbol{(} 
\Tr [U^{\dagger} (\varphi)] \boldsymbol{)} U (\varphi)$ is 
parallel to the initial unitary $U(0)=\hat{1}$, i.e.,  
$\Tr [\overline{U} (\varphi)] \geq 0$. We can identify this with Eq. 
(\ref{eq:finalfinal}). Hence,  $\Lambda^{{\textrm{u}}}$ is the 
gauge-invariant gluing between the two unitary end point channels 
$\mathcal{E}_0^{{\textrm{u}}} = \mathcal{I}$ and 
$\mathcal{E}_{\varphi}^{{\textrm{u}}}$. Furthermore, the 
phase factor $\gamma_{{\textrm{ch}}} (C)$ is the corresponding 
gluing matrix with respect to the parallel Kraus representations 
$U(0)=\hat{1}$  and  $\overline{U} (\varphi)$ of the end point channels.

In the experiment in Refs. \cite{rauch75,werner75} the input state to
the interferometer was $|0\rangle\langle 0|\otimes\frac{1}{2} \hat{1}$
(unpolarized neutrons).  If we assume that the beam splitters can be
represented by Hadamard gates, the passage through the first
beam splitter results in the state $\sigma = \frac{1}{2}
(\ket{0}+\ket{1}) (\bra{0}+\bra{1})
\otimes \frac{1}{2} \hat{1}$. After the application of the channel 
$\Lambda^{{\textrm{u}}}$ on this state the particle passes through 
a second beam splitter, and the probability to find the neutron in 
the $0$ beam, say, is
\begin{eqnarray} 
p(\varphi) = \frac{1}{2} \Big( 1 + \big| \Tr [U(\varphi)] \big| 
\cos[ \arg \gamma_{{\textrm{ch}}}(C) ] \Big) ,  
\end{eqnarray}  
where we have used that $\Tr [\overline{U} (\varphi)] = \big| 
\Tr [U (\varphi)] \big|$. 
We see that $p(\varphi)$ has period $4\pi$ in $\varphi$. Since the
visibility factor $| \Tr U(\varphi) | = | \cos \frac{\varphi}{2} |$ is
$2\pi$ periodic in $\varphi$, the observed periodicity of the
interference oscillations must originate solely from the $4\pi$
periodicity of the channel holonomy $\gamma_{{\textrm{ch}}} [C]$. This
concludes our demonstration that Refs. \cite{rauch75,werner75} can be
viewed as experimental realizations of a channel holonomy and 
its $4\pi$ periodicity.

Finally, we wish to point out that the holonomy for this family of 
unitary channels could also be measured. It requires the following 
modification of the setup in Refs. \cite{rauch75,werner75}: add 
a U(1) shift $e^{i\chi}$ to the $1$ beam and maximize the output 
detection probability $p$ by varying $\chi$. A direct calculation 
yields 
\begin{eqnarray}
p(\chi) = \frac{1}{2} \Big( 1 + \big| \Tr [U(\varphi)] \big| 
\cos[ \chi - \arg \gamma_{{\textrm{ch}}}(C) ] \Big) , 
\end{eqnarray}
which is maximal when $\chi = \arg \gamma_{{\textrm{ch}}}(C)$. 

\section{Conclusions}
Investigations into quantum holonomy have yielded a unifying
understanding of the geometry of some basic structures of quantum
systems. These efforts have concerned not only the geometry of quantum
states themselves, but also how the twisting of subspaces, realized,
e.g., as eigenspaces of some adiabatically varying Hamiltonian, can be
used to manipulate quantum states in a robust manner.

In this paper, we have extended the notion of holonomy to sequences of
quantum channels. The proposed quantity transforms as a holonomy under
change of Kraus representations of the channels. We
have shown that the channel holonomy is related to the Uhlmann
holonomy \cite{uhlmann86} for sequences of density operators
constructed from the Jamio{\l}kowski isomorphism
\cite{jamiolkowski72}. We have delineated parallel transport and
concomitant gauge potential for smooth families of channels.

In addition to the relation to the Uhlmann holonomy, we have found
some results that connect to other known quantum holonomies. For
smooth sequences of unitary channels, the channel holonomy reduces to
the phase shift in Ref. \cite{sjoqvist00} for unpolarized particles in
an interferometer. Furthermore, we have analyzed a class of channels
that have a direct relation to the subspace holonomies in
Ref. \cite{kult06}. For smooth families of such ``holonomic'' 
channels, the channel holonomy is completely determined by
the trace of the holonomies associated with paths in the space of
subspaces (i.e., the Wilson loops if these paths are closed).

We have demonstrated a physical realization of the channel holonomy in
an interferometric setting, based on the idea of ``gluings'' of
channels \cite{aberg04a,aberg04b}. The realization consists of a gauge
invariant object related to the channel holonomy and a prescription
for how this object can be used to extract the channel holonomy
experimentally.  Using this idea, we have been able to demonstrate
that the neutron interferometer tests in Refs. \cite{rauch75,werner75}
of the $4 \pi$ spinor symmetry can alternatively be interpreted in
terms of the holonomy for unitary channels. Thus, one may be tempted
to say that a particular realization of the channel holonomy already
exists.

To avoid technical complications, we have consistently made
simplifying assumptions about the nature of the channels. We have
restricted the analysis to sequences of channels with fixed number of
linearly independent Kraus operators and we have assumed that all
relevant matrices and operators have a well-defined inverse.
If these assumptions are relaxed, one could consider analogues
to the admissible sequences of density operator as considered
by Uhlmann \cite{uhlmann86}, or to the partial holonomies
described in Refs. \cite{kult06,sjoqvist06}.

Let us end by a remark concerning the potential relevance of this work
to holonomic quantum computation \cite{zanardi99,pachos00}. The key
point with this type of quantum computation is that it is believed to
be resilient to certain errors, such as those induced by open-system
effects. Thus, in order to examine the resilience of holonomic quantum
computation, it becomes important to have a useful notion of geometric
phase or holonomy for open quantum systems. This has been addressed 
from different perspectives in several recent papers 
\cite{ericsson03,peixoto03,carollo03,marzlin04,sarandy06,bassi06,goto07}. 
Since open-system evolution may be described by quantum channels, it
seems reasonable that the proposed channel holonomy, or some
generalization of it allowing for variable Kraus number, might, in
some way or another, play a role in the analysis of the resilience
of holonomic quantum computation to open-system effects.

\section*{ACKNOWLEDGMENTS}
J.{\AA}. wishes to thank the Swedish Research Council for financial 
support and the Centre for Quantum Computation at DAMTP, Cambridge, 
for hospitality. E.S. acknowledges financial support from the 
Swedish Research Council. The work by J.{\AA}. was supported by the 
European Union through the Integrated Project QAP (IST-3-015848), 
SCALA (CT-015714), SECOQC and the QIP IRC (GR/S821176/01).

\section*{APPENDIX} 
Consider channels $\mathcal{E}$ and $\mathcal{F}$ both with Kraus
number $K$. Any choice of Kraus operators for this pair of channels
span $K$-dimensional subspaces $\mathcal{L}_{\mathcal{E}}$ and
$\mathcal{L}_{\mathcal{F}}$, respectively, of the $D^2$-dimensional
space of linear operators acting on the $D$-dimensional state
space. We show that a necessary and sufficient criterion for any
$\boldsymbol{T}$ corresponding to $\mathcal{E}$ and $\mathcal{F}$ to
have rank less than $K$ is that $\mathcal{L}_{\mathcal{E}}$ and
$\mathcal{L}_{\mathcal{F}}$ are not fully overlapping.
 
Assume $\mathcal{L}_{\mathcal{E}}$ and $\mathcal{L}_{\mathcal{F}}$ 
are partially overlapping. Then, one can choose a Kraus representation 
$\{ F_k \}_k$ for $\mathcal{F}$ so that there exists a $F_{k'}$ in 
this set lying in the orthogonal complement to $\mathcal{L}_{\mathcal{E}}$. 
It follows that $\boldsymbol{T}_{k'l} = \Tr (F_{k'}^{\dagger} E_l) = 0$, 
$\forall l$, and thus the rank of $\boldsymbol{T}$ is less than $K$.  

Conversely, if the rank of $\boldsymbol{T}$ is less than $K$, 
then $\boldsymbol{T}$ has at least one singular value that is zero. 
Consider the singular value decomposition $\boldsymbol{T}=
\boldsymbol{U}\boldsymbol{D}\boldsymbol{V}$ ($\boldsymbol{D}$ 
diagonal and $\boldsymbol{U},\boldsymbol{V}$ unitary). Assume 
$\boldsymbol{D}_{k'k'}=0$ and consider the transformation 
$F_k \rightarrow \widetilde{F}_k = \sum_l F_l \boldsymbol{U}_{lk}$.  
This results in $\boldsymbol{T} \rightarrow \widetilde{\boldsymbol{T}}=
\boldsymbol{D}\boldsymbol{V}$, which implies 
$\widetilde{\boldsymbol{T}}_{k'l} = \Tr (F_{k'}^{\dagger} E_l) = 0$, 
$\forall l$. Thus, $\widetilde{F}_{k'}$ lies in the orthogonal complement 
to $\mathcal{L}_{\mathcal{E}}$.  

As an illustration, consider the following representations  
\begin{eqnarray}
\{ E_0,E_1 \} & = & 
\left\{ \sqrt{1-p_e}) \hat{1},\sqrt{p_e} \sigma_z \right\} , 
\nonumber \\ 
\{ F_0,F_1 \} & = & 
\left\{ \sqrt{1-p_f} \hat{1} , \sqrt{p_f} \sigma_x \right\} , 
\nonumber \\ 
(G_0,G_1) & = & 
\left\{ \frac{1}{2} \big( 1+\sqrt{1-p_g} \big) \hat{1} \right.  
\nonumber \\ 
 & & \left. + \frac{1}{2} \big( 1-\sqrt{1-p_g} \big) \sigma_z ,
\frac{1}{2}\sqrt{p_g} \sigma_+ \right\}   
\end{eqnarray}
of the phase flip ($\mathcal{E}$), bit flip ($\mathcal{F}$), and  
amplitude damping ($\mathcal{G}$) channels for a qubit ($D=2$).  
Here, the $\sigma$'s are the standard Pauli operators (with $\sigma_+ = 
\sigma_x + i\sigma_y$) and $p_e,p_f,p_g \in [0,1]$. Clearly, 
$K(\mathcal{E}) = K(\mathcal{F}) = K(\mathcal{G}) = K = 2$. By
inspection, we see that $F_1$ and $G_1$ both lie in the orthogonal
complement to $\mathcal{L}_{\mathcal{E}}$. Thus, both
$\mathcal{L}_{\mathcal{F}}$ and $\mathcal{L}_{\mathcal{G}}$ overlap
partially with $\mathcal{L}_{\mathcal{E}}$ and the corresponding
$\boldsymbol{T}$ matrices have rank less than $K=2$. Explicitly, one
obtains
\begin{eqnarray} 
\boldsymbol{T}_{F,E} & = & 
\begin{pmatrix}
2 \sqrt{(1-p_e)(1-p_f)} & 0 \\
0 & 0 &
\end{pmatrix} , 
\nonumber \\ 
\boldsymbol{T}_{E,G} & = & 
\begin{pmatrix}
\sqrt{1-p_e} \big( 1+\sqrt{1-p_g} \big) & 
0 \\
\sqrt{p_e} \big( 1-\sqrt{1-p_g} \big) & 0 &
\end{pmatrix} . 
\end{eqnarray}  
Clearly, the rank of $\boldsymbol{T}_{F,E}$ is $1$ if 
$p_e,p_f \neq 1$ and $0$ ($\mathcal{L}_{\mathcal{E}}$ and 
$\mathcal{L}_{\mathcal{F}}$ orthogonal) if $p_e = 1$ or 
$p_f = 1$. The rank of $\boldsymbol{T}_{G,E}$ is $0$ 
if $p_e = 1$ and $p_g = 0$, and $1$ otherwise.   
On the other hand, 
\begin{eqnarray} 
\boldsymbol{T}_{G,F} =  \begin{pmatrix}
\sqrt{1-p_f} \big( 1+\sqrt{1-p_g} \big) & 0 \\
0 & \sqrt{p_g} \sqrt{1-p_f} &
\end{pmatrix} , 
\end{eqnarray}  
which has rank $K=2$ if $p_f \neq 1$ and $p_g \neq 0$.


\begin{thebibliography}{99}
\bibitem{berry84} M. V. Berry, 
Proc. R. Soc. London A {\bf 392}, 45 (1984).  
\bibitem{wilczek84} F. Wilczek and A. Zee,
Phys. Rev. Lett. {\bf 52}, 2111 (1984).
\bibitem{aharonov87} Y. Aharonov and J. Anandan,
Phys. Rev. Lett. {\bf 58}, 1593 (1987).
\bibitem{samuel88} J. Samuel and R. Bhandari,
Phys. Rev. Lett. {\bf 60}, 2339 (1988).
\bibitem{uhlmann86} A. Uhlmann, 
Rep. Math. Phys. {\bf 24}, 229 (1986). 
\bibitem{anandan89} J. Anandan and A. Pines,
Phys. Lett. A {\bf 141}, 335 (1989). 
\bibitem{anandan88} J. Anandan,
Phys. Lett. A {\bf 133}, 171 (1988). 
\bibitem{wu05} L.-A. Wu, P. Zanardi, and D. A. Lidar, 
Phys. Rev. Lett. {\bf 95}, 130501 (2005).
\bibitem{carollo06} A. Carollo, M. Fran\c{c}a Santos, and V. Vedral, 
Phys. Rev. Lett. {\bf 96}, 020403 (2006).
\bibitem{zanardi99} P. Zanardi and M. Rasetti, 
Phys. Lett. A {\bf 264}, 94 (1999). 
\bibitem{pachos00} J. Pachos, P. Zanardi, and M. Rasetti,
Phys. Rev. A {\bf 61}, 010305(R) (2000).
\bibitem{jamiolkowski72} A. Jamio{\l}kowski,
Rep. Math. Phys. {\bf 3}, 275 (1972). 
\bibitem{aberg04a} J. {\AA}berg, 
Ann. Phys. (N.Y.) {\bf 313}, 326 (2004). 
\bibitem{aberg04b} J. {\AA}berg, 
Phys. Rev. A {\bf 70}, 012103 (2004).
\bibitem{rauch75} H. Rauch, A. Zeilinger, G. Badurek,
A. Wilfing, W. Bauspiess, and U. Bonse,  
Phys. Lett. A {\bf 54}, 425 (1975).     
\bibitem{werner75} S. A. Werner, R. Colella, A. W. Overhauser, 
and C. F. Eagen, 
Phys. Rev. Lett. {\bf 35}, 1053 (1975).   
\bibitem{kraus83} K. Kraus, 
{\it States, Effects, and Operations} (Springer-Verlag, Berlin, 1983). 
\bibitem{remark1} More generally, the summation range on the 
right-hand side of Eq. (\ref{eq:kraustransform}) can be larger 
than the Kraus number $K(\mathcal{F})$, if we demand that 
$\boldsymbol{U}$ is a partial isometry. In such cases, the 
resulting operators $\{ \widetilde{F}_k \}_k$ become linearly dependent. 
\bibitem{remark2} Consider the set of channels with Kraus 
number $K$, and two elements $\mathcal{E}$ and $\mathcal{F}$ from this
set. Consider moreover $d(\mathcal{E},\mathcal{F}) =
\inf_{\{F_{k}\}_{k}}\sqrt{\sum_{k}||E_{k}-F_{k}||^{2}}$, where
$\{E_{k}\}_{k}$ is a linearly independent Kraus representation of
$\mathcal{E}$, and where the infimum is taken over all linearly
independent Kraus representations $\{F_{k}\}_{k}$ of $\mathcal{F}$. On
a set of channels with a fixed Kraus number, $d(\cdot,\cdot)$ is a
proper metric, i.e., $d(\mathcal{E},\mathcal{F}) =
d(\mathcal{F},\mathcal{E})$, $d(\mathcal{E},\mathcal{F})\geq 0$
with equality if and only if $\mathcal{E}=\mathcal{F}$, and it
satisfies the triangle inequality $d(\mathcal{E},\mathcal{F}) \leq
d(\mathcal{E},\mathcal{G}) + d(\mathcal{G},\mathcal{F})$.
\bibitem{comment} The non-Abelian open path holonomy in Ref. 
\cite{kult06} and its special case in Ref. \cite{wilczek84} for 
closed paths provide examples of an holonomy
in the form of a gauge-covariant unitary matrix $\boldsymbol{U}_{g}$
that represents a gauge-invariant partial isometry $\Gamma$ via the
relation $\Gamma = \sum_{kl}[\boldsymbol{U}_{g}]_{kl} 
|\overline{a}_{k}(1)\rangle\langle a_{l}(0)|$ (see Eq. (12) 
in Ref. \cite{kult06}). 
\bibitem{kult06} D. Kult, J. {\AA}berg, and E. Sj\"oqvist,  
Phys. Rev. A {\bf 74}, 022106 (2006). 
\bibitem{remark3} Extending the sequence accommodates the 
multiplication by $\Phi(\boldsymbol{T}_{1,N})$ in the definition 
of the channel holonomy in Eq.~(\ref{eq:kinholonomy}). 
\bibitem{uhlmann91} A. Uhlmann, 
Lett. Math. Phys. {\bf 21}, 229 (1991). 
\bibitem{bhatia97} R. Bhatia, 
{\it Matrix Analysis}, Graduate texts in mathematics, vol. 169
(Springer, Berlin, 1997).  
\bibitem{yang54} C. N. Yang and R. L. Mills, 
Phys. Rev. {\bf 96}, 191 (1954).  
\bibitem{dittmann92} J. Dittmann and G. Rudolph, 
J. Math. Phys. {\bf 33}, 4148 (1992). 
\bibitem{sjoqvist00} E. Sj\"oqvist, A. K. Pati, A. Ekert, 
J. S. Anandan, M. Ericsson, D. K. L. Oi, and V. Vedral, 
Phys. Rev. Lett. {\bf 85}, 2845 (2000). 
\bibitem{zee88} A. Zee, 
Phys. Rev. A {\bf 38}, 1 (1988). 
\bibitem{remark4} Note that the holonomy for unitary channels 
can be more general phase factors for higher $n$ since 
$\Tr$\big(SU(n$>$2)\big) is not restricted to real values. 
\bibitem{note} One may note that the gauge transformation of 
the Kraus representations in Eq. (\ref{eq:kraustransform}) 
relates to the gauge transformation $\mathcal{U}\rightarrow 
(\hat{1} \otimes U)\mathcal{U}$ according to  
$\langle a_{k}|U|a_{l}\rangle = [\boldsymbol{U}^{{\textrm{T}}}]_{kl}$, 
where ${{\textrm{T}}}$ denotes transpose with respect to some 
orthonormal basis $\{|a_{k}\rangle\}_{k=1}^{K}$ of $\mathcal{H}_a$. 
\bibitem{oi03} D. K. L. Oi, 
Phys. Rev. Lett. {\bf 91}, 067902 (2003).
\bibitem{oi06} D. K. L. Oi and J. {\AA}berg, 
Phys. Rev. Lett. {\bf 97}, 220404 (2006).
\bibitem{aberg07} J. {\AA}berg, D. Kult, E. Sj\"oqvist, and 
D. K. L. Oi, 
Phys. Rev. A, {\bf 75}, 032106 (2007).
\bibitem{bernstein67} H. J. Bernstein, 
Phys. Rev. Lett. {\bf 18}, 1102 (1967).
\bibitem{aharonov67} Y. Aharonov and L. Susskind, 
Phys. Rev. {\bf 158}, 1237 (1967).
\bibitem{sjoqvist06} E. Sj\"oqvist, D. Kult, and J. {\AA}berg, 
Phys. Rev. A {\bf 74}, 062101 (2006). 
\bibitem{ericsson03} M. Ericsson, E. Sj\"oqvist, J. Br\"annlund, 
D. K. L. Oi, and A. K. Pati, 
Phys. Rev. A {\bf 67}, 020101(R) (2003).   
\bibitem{peixoto03} J. G. Peixoto de Faria, A. F. R. de Toledo Piza,   
and M. C. Nemes, 
Europhys. Lett. {\bf 62}, 782 (2003).   
\bibitem{carollo03} A. Carollo, I. Fuentes-Guridi, 
M. Fran\c{c}a Santos, and V. Vedral,  
Phys. Rev. Lett. {\bf 90} 160402 (2003).   
\bibitem{marzlin04} K.-P. Marzlin, S. Ghose, and B. C. Sanders, 
Phys. Rev. Lett. {\bf 93}, 260402 (2004). 
\bibitem{sarandy06} M. S. Sarandy and D. A. Lidar,  
Phys. Rev. A {\bf 73}, 062101 (2006).   
\bibitem{bassi06} A. Bassi and E. Ippoliti, 
Phys. Rev. A {\bf 73}, 062104 (2006). 
\bibitem{goto07} H. Goto and K. Ichimura, 
Phys. Rev. A {\bf 76}, 012120 (2007).  
\end{thebibliography}
\end{document}